\documentclass[acmlarge, authorversion,nonacm]{acmart}
\usepackage{graphicx} % Required for inserting images
\usepackage{amsmath}
\usepackage{balance}       % to better equalize the last page
\usepackage{graphics}   
\usepackage{graphicx}
\usepackage[T1]{fontenc}   % for umlauts and other diaeresis
\usepackage{lipsum}
\usepackage{color}
\usepackage{colortbl}
\usepackage{booktabs}
\usepackage{textcomp}
\usepackage{microtype}        % Improved Tracking and Kerning
\usepackage{ccicons}          % Cite your images correctly!
\usepackage{multirow}  % for the tables
\usepackage{color,soul}  % For highlighting
\usepackage{subcaption}  % subfigure
\usepackage{listings}  % inline code snippets
\usepackage{todonotes}
\usepackage{multirow}

\usepackage{array}
\newcolumntype{P}[1]{>{\centering\arraybackslash}p{#1}}
 
\usepackage[capitalise]{cleveref}

\def\etal{\emph{et al.\ }}

\usepackage{arydshln}
\usepackage{fancyvrb}
\usepackage[frozencache=true, cachedir=minted-cache]{minted}

\definecolor{lightgraybackground}{rgb}{0.945, 0.949, 0.957}

\title{Past, Present, and Future of Sensor-based \underline{H}uman \underline{A}ctivity \underline{R}ecognition using Wearables:  
A Surveying Tutorial on a Still Challenging Task}

\begin{document}

\author{Harish Haresamudram}
\affiliation{%
  \institution{School of Interactive Computing, College of Computing, Georgia Institute of Technology}
  \streetaddress{85 5th Street N.W.}
  \city{Atlanta}
  \state{GA}
  \postcode{30332}
  \country{USA}}
\email{harishkashyap@gatech.edu}

\author{Chi Ian Tang}
%\authornote{Both authors contributed equally to this research.}
\email{cit27@cl.cam.ac.uk}
%\orcid{1234-5678-9012}
%\author{G.K.M. Tobin}
%\authornotemark[1]
%\email{webmaster@marysville-ohio.com}
\affiliation{%
  \institution{Nokia Bell Labs}
  %\streetaddress{P.O. Box 1212}
  \city{Cambridge}
  %\state{Ohio}
  \country{UK}
  %\postcode{43017-6221}
}

\author{Sungho Suh}
\email{sungho.suh@dfki.de}
\affiliation{%
  \institution{DFKI}
  \streetaddress{Trippstadter Str. 122}
  \postcode{67663}
  \city{Kaiserslautern}
  \country{Germany}
}
\affiliation{%
  \institution{RPTU Kaiserslautern-Landau}
  % \streetaddress{Trippstadter Str. 122}
  % \postcode{67663}
  \city{Kaiserslautern}
  \country{Germany}
}

\author{Paul Lukowicz}
\email{paul.lukowicz@dfki.de}
\affiliation{%
  \institution{DFKI}
  \streetaddress{Trippstadter Str. 122}
  \postcode{67663}
  \city{Kaiserslautern}
  \country{Germany}
}
\affiliation{%
  \institution{RPTU Kaiserslautern-Landau}
  % \streetaddress{Trippstadter Str. 122}
  % \postcode{67663}
  \city{Kaiserslautern}
  \country{Germany}
}

\author{Thomas Pl{\"o}tz}
\email{thomas.ploetz@gatech.edu}
\affiliation{%
  \institution{School of Interactive Computing, College of Computing, Georgia Institute of Technology}
  \streetaddress{85 5th Street N.W.}
  \city{Atlanta}
  \state{GA}
  \postcode{30332}
  \country{USA}}

\renewcommand{\shortauthors}{Haresamudram et al.}
\renewcommand{\shorttitle}{Past, Present, and Future of Sensor-Based Human Activity Recognition using Wearables}

%%
%% The abstract is a short summary of the work to be presented in the
%% article.
\begin{abstract}
In the many years since the inception of wearable sensor-based Human Activity Recognition (HAR), a wide variety of methods have been introduced and evaluated for their ability to recognize activities. 
Substantial gains have been made since the days of hand-crafting heuristics as features, yet, progress has seemingly stalled on many popular benchmarks, with performance falling short of what may be considered `sufficient'--despite the increase in computational power and scale of sensor data, as well as rising complexity in techniques being employed. 
The HAR community approaches a new paradigm shift, this time incorporating world knowledge from foundational models.
In this paper, we take stock of sensor-based HAR -- surveying it from its beginnings to the current state of the field, and charting its future.
This is accompanied by a hands-on tutorial, through which we guide practitioners in developing HAR systems for real-world application scenarios. 
We provide a compendium for novices and experts alike, of methods that aim at finally solving the activity recognition problem. 
\end{abstract}

%
% The code below should be generated by the tool at
% http://dl.acm.org/ccs.cfm
% Please copy and paste the code instead of the example below. 
%

%
% End generated code
%

\maketitle

\section{Introduction}
\label{sec:introduction}

With a history of thirty years or so of very active research and development in human activity recognition (HAR), one would expect that the problem of automatically recognizing what a person is doing (and when) should be solved by now, i.e., that sensor-based HAR using wearables is now "good enough" to have become a commodity and widely accepted. 
In fact, many commercially available wearables such as smart watches include--variants of--HAR as a central service element and even selling point.
It seems appealing to end users to automatically track the steps they have taken during a day, count the repetitions of a free weights workout, analyze their sleep, or even estimate the calories they have burnt.
HAR based on the analysis of body-worn movement sensors serves as the algorithmic foundation for many of these tasks, albeit at times with questionable accuracy \cite{duking2024smartwatch}.

Yet, activity recognition goes beyond such ``low hanging fruits'' and the research community is now attempting more detailed activity assessments such as longitudinal health monitoring \cite{sangeethalakshmi2023patient} including change detection, detailed sports tracking and coaching \cite{margarito2015user,zhou2022quali,singh2024novel,ladha2013climbax,kranz2013mobile},  or quality control and process tracking in manufacturing \cite{tao2018worker,suh2023worker,bello2024tsak} to name but a few.
When tackling such non-trivial activity recognition problems it quickly becomes clear that HAR is still far from being a commodity with recognition performance on challenging benchmark tasks such as the Opportunity challenge \cite{chavarriaga2013opportunity} stagnating now for more than a decade.
Yet, not all is lost as recent breakthroughs in the broader field of Artificial Intelligence have been creatively adopted and adapted by the HAR research community, leading to tailored solutions that substantially push the state-of-the-art.

With that, it is time to take inventory of where the field stands and to summarize how to tackle HAR in practical applications.
This is what this paper sets out to achieve: To provide a survey of the past, present, and future of sensor-based Human Activity Recognition using wearables -- and to compile a tutorial for practitioners on how to approach HAR in practical, real-world applications.
This paper is based on the collective experience and expertise of the authors who have been working in the field for decades, and on a series of tutorials that were held at the annual flagship conferences of the field.
This tutorial is accompanied by a code-base and a set of experiments (along with instructions) that will allow the interested reader to not only follow along with the explanations given here but also to integrate state-of-the-art HAR techniques into their own practical applications.

Arguably, the most pressing issue for HAR research remains the lack of labeled sample data -- which often leads to poor generalization capabilities of activity recognition systems overall. 
Much of contemporary research aims to overcome this roadblock through: 
\textit{i)} representation learning; 
\textit{ii)} multi- or cross-modality learning approaches including generative, augmentative, and simulation methods; or
\textit{iii)} foundational models that aim for incorporating world contextual knowledge into the specific HAR tasks -- or combinations of these three categories.
Accordingly, our focus is on representations and modeling techniques that exploit multiple modalities in effective ways.

The goal of this paper is two-fold:
We survey the past and present of relevant HAR research in the field of wearable and ubiquitous computing. 
By doing so we contextualize our hands-on tutorial for practitioners who aim to develop practical HAR applications thereby tackling challenging scenarios that typically require more than mere "out of the box" deployment of existing methods. 

\subsection{Relation to Existing Surveys and Tutorials}
Previous surveys and tutorials have reviewed different aspects of HAR thereby reflecting the increasing attention and efforts dedicated to the field. Our paper extends and complements these previous papers by focusing on recent developments and specifically targeting challenging, non-trivial real-world applications of HAR.

Lara \& Labrador \cite{lara2012survey} provided an extensive review of early works in wearable-based HAR systems. 
Key design issues were discussed in the survey, including energy consumption, sensor placement, and flexibility. 
Bulling et al.\ \cite{bulling2014tutorial} compiled a seminal tutorial paper that explicitly captures the classical, pre-Deep-Learning era of HAR with wearable sensors.
It defined the five-stage Activity Recognition Chain (ARC) as the de-facto standard for the field for many years. 
While this tutorial remains highly relevant, its primary focus was on classical machine learning techniques, leaving current trends such as the rise of deep learning models, representation learning with unlabeled data, and more recently, learning from multi-modal data under-explored.

This is where our surveying tutorial comes into play by providing an up-to-date overview of the field and hands-on explanations for practitioners on how to tackle challenging, real-world HAR problems.

\subsection{Scope and Organization}
This tutorial targets human activity recognition through body-worn movement sensors and machine learning-based sensor data analysis that draws from classic signal processing as well as contemporary Artificial Intelligence.
The main focus lies on how to represent activity data combined with questions related to effective modeling.

In Sec.\ \ref{sec:history} we provide a concise survey of the field's history from its origins with handcrafted features and classical ML-based classifiers, to contemporary end-to-end learning. 
We then focus on the most pressing issue of how rich, learned representations push the field (Sec.\ 
 \ref{sec:ssl}) specifically covering the successful adoption of self-supervised learning (SSL) methods and aspects of multi-modality.
Sec.\ \ref{sec:generation} focuses on generating / augmenting sample data using contemporary AI methods, and in Sec.\ \ref{sec:future} we discuss the prospects of foundational models for the field.
We conclude with a discussion and provide links and instructions for the accompanying code-base.

\section{History: from hand crafted features to DNNs}
\label{sec:history}
\begin{figure}[t]
    \centering
    \includegraphics[width=1\textwidth]{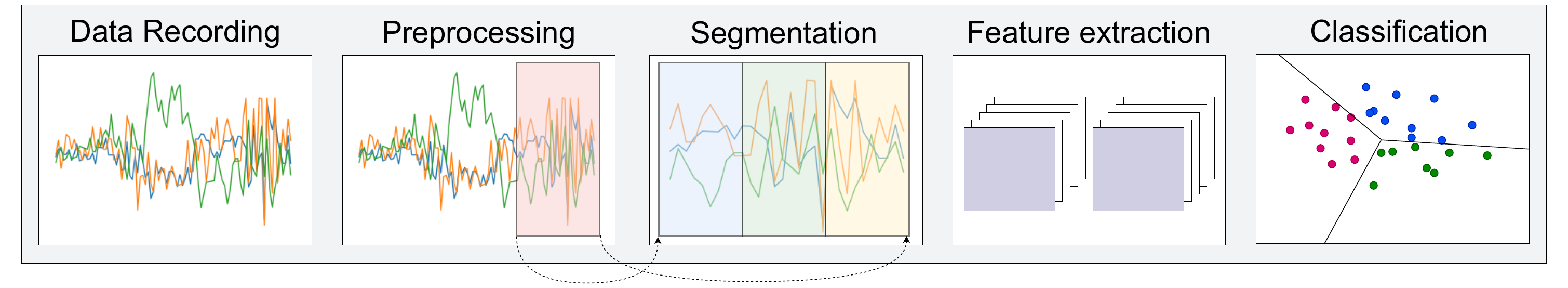}
    \caption{
    The Activity Recognition Chain as summarized in \cite{bulling2014tutorial}. 
     }
    \label{fig:arc}
\end{figure}

Sensor-based human activity recognition (HAR) corresponds to the automatic classification of sensor data into activities of interest (or the null class).
Traditionally it was performed in five steps (Fig.\  \ref{fig:arc}) through the Activity Recognition Chain (ARC) as summarized by Bulling \etal \cite{bulling2014tutorial}:
\emph{(i) data collection:} where (body-worn) movement sensors such as inertial measurement units (IMU) are used to directly record activity data from wearers; 
\emph{(ii) pre-processing:} which includes filtering, denoising, normalization, etc.\ preparing the data at a signal processing level for subsequent activity recognition;
\emph{(iii) segmentation:} where a sliding window approach is used to aggregate and cut out contiguous segments of sensor data from the stream of readings; 
\emph{(iv) feature extraction:} where compact, meaningful representations--features--are extracted from aforementioned windows of (preprocessed)  sensor data; and
\emph{(v) classification:} where features are classified into activities, typically employing machine learning methods.

\subsection{Feature Engineering and the Activity Recognition Chain (ARC)}
Traditionally, the ARC employed handcrafted, statistical \cite{figo2010preprocessing,plotz2011activity}  and distribution-based features \cite{hammerla2013preserving,kwon2018adding}, or techniques such as Principle Component Analysis (PCA) \cite{plotz2011feature}.
The goal was to effectively represent the movement present in windows of sensor data into vectors that are (hopefully) useful for recognizing activities. 
The process required substantial manual effort to discover and recognize through trial-and-error, which features are effective for HAR.
In addition to this substantial manual effort, most of such handcrafted, typically heuristics-driven features did not generalize well across application domains.
In response, researchers focused specifically on automatically deriving--learning--rich and especially generalizable feature representations and integrated these into the overall ARC \cite{plotz2011feature}. 
 
Activity classification itself as part of the ARC was performed through ``classical'' machine learning approaches such as Support Vector Machines (SVM), logistic regression, k Nearest Neighbors (kNNs), or Random Forest (RF) classifiers -- each processing feature vectors of individual sensor data windows.\footnote{Such a sliding window based approach--while functional in general--comes with at least two problems: subsequent, often overlapping, windows of sensor readings are not \textit{i.i.d.} and thus care needs to be taken during model training and evaluation \cite{hammerla2015let}, and using the same temporal context for both feature extraction and activity modeling limits applicability \cite{hiremath2021role,li2018specialized}}
Overall, feature extraction and classification in the ARC, and its many variants, are not directly coupled making it challenging to derive useful features in a systematic manner.

\paragraph{Statistical Features:} These representations essentially resemble heuristics on capturing certain statistical aspects of the underlying sensor signals \cite{haresamudram2019role}, including: 
\emph{(i)} DC mean of the signal; 
\emph{(ii)} its variance; 
\emph{(iii)} the correlation (between channels);  
\emph{(iv)} signal energy; and 
\emph{(v)} frequency-domain entropy, to name but a few examples. 
The DC mean comprises the averaged sensor data in the window, whereas the variance characterizes the stability of the signal. 
Energy captures the periodicity of the signal and the frequency domain entropy helps discriminate between activities of similar energy. 
The correlation is computed between all pairwise combinations of axes and captures the correlation between different axes.
While such measures capture general (statistical) features of movement signals well (enough), they have no actual connection to the actual classification domain in the sense that they are too generic for robust and targeted HAR.
    
Open-source libraries exist that facilitate for computing of hundreds of such heuristic features and thus lower the bar for practitioners entering the field (e.g., \texttt{tsfresh} \cite{christ2018time}).
The resulting--often very--high dimensional features can be effective at recognizing activities, yet explicit post-processing is needed for practical applicability to combat the high dimensionality issues of subsequent model training.

\paragraph{Distribution-based Features:} As proposed by Hammerla \etal \cite{hammerla2013preserving} and later refined by Kwon \etal \cite{kwon2018adding}, a meaningful alternative to heuristic, handcrafted features are representations that directly cover relevant aspects of the distributions underlying a window of sensor data as it is processed by the ARC.
Specifically for a compact representation the inverse of the Empirical Cumulative Distribution Function (ECDF) is computed and its (subset of) quantiles are then used as features for HAR.

As shown by Haresamudram \etal \cite{haresamudram2019role}, these features are not computationally intensive and can be computed on the fly on many wearable devices, even those with severe resource limitations.
Furthermore, they have been used in HAR systems that recognize standard sets of activities (e.g., running, walking, sitting, standing, etc.) with reasonably high accuracy.
Yet, the applicability of the ARC, including its refinements with regard to all of its five components and specifically with the optimizations of feature representations, remains limited to coarse-grained activity recognition.
More detailed activity assessments, such as the Opportunity challenge that aims at recognizing complex, less repetitive household activities \cite{chavarriaga2013opportunity} remain challenging.

\subsection{End-to-End Learning Based Approaches}
With the availability of very large, labeled datasets on the internet and the virtual disappearance of computational constraints through the introduction of cloud and GPU computing in the early 2010s, many ML application domains shifted to modifying the way artificial neural networks as classification backends were configured, trained, and used \cite{lecun2015deep}, i.e., Deep Learning (DL) was introduced.
Instead of shallow model architectures with typically one hidden layer and a moderate number of neurons--all owed to the substantial, former restrictions on available sample data as well as computational resources--new model architectures were introduced that contained dozens of hidden layers and very large numbers of neurons.
It was shown that such models--if trained properly--outperform conventional ML models, including neural networks, by substantial margins \cite{krizhevsky2012imagenet}. 

This led to many communities quickly adopting Deep Learning, leading to significant performance gains -- if sufficient amounts of labeled training data were available. 
The core transformation introduced with DL was a reduced emphasis on designing and handcrafting features but rather \textit{learning} representations in an \textit{end-to-end} manner as part of the overall modeling and training procedure.
Whilst the HAR community does not have access to labeled datasets that are remotely comparable in size to those that are standard in, for example, the computer vision (CV) or the natural language processing (NLP) communities, the idea of learning representations gained attraction here, too.
Deeper, yet not as deep as in CV or NLP, model architectures such as the DeepConvLSTM \cite{ordonez2016deep} were introduced specifically eliminating the explicit feature design phase but rather utilizing proven feature learners such as convolutional blocks into the model architectures, complemented by explicit sequential modeling parts in form of LSTM blocks \cite{hochreiter1997long} or later ensembles thereof \cite{guan2017ensembles}.
Mainly focusing on the implicit representation learning aspect the model architectures had to remain less complex though (compared to CV and NLP models) due to the lack of large enough labeled training sets.

Yet, DL-based end-to-end training rose to prominence for wearables-based HAR. 
For example, Zeng \etal \cite{zeng2014convolutional} presented a convolutional network for recognizing activities.
Various types of layers, including fully connected, convolutional, and recurrent networks were studied by Hammerla \etal \cite{hammerla2016deep}.
A combination of convolutional and recurrent layers was proposed in DeepConvLSTM \cite{ordonez2016deep}, which, even today, remains a strong baseline and is commonly utilized in contemporary HAR works \cite{haresamudram2022assessing, thukral2023cross}.
Bock \etal \cite{bock2021improving}, however, found that using a shallower LSTM in the DeepConvLSTM setup is better for many HAR datasets.
More convolutional architectures have since been evaluated, typically involving deeper networks and residual connections \cite{kwon2021complex, shao2023convboost, yuan2024self}.

Going beyond, attention models were explored, with the ability to automatically `attend to' relevant parts of input data, typically using the output of recurrent networks.
Temporal attention was applied by Murahari \etal \cite{murahari2018attention} whereas continuous sensor and temporal attention were evaluated by Zeng \etal \cite{zeng2018understanding}, both leading to increased performance. 
In contrast, TinyHAR \cite{zhou2022tinyhar} also utilizes attention modules, but focuses on being lightweight for deployments. 
More recently, the self-attention mechanism introduced in the Transformer paper \cite{vaswani2017attention} has also been successfully applied for sensor-based HAR \cite{mahmud2020human, gao2021danhar, cao2023human, gao2023mmtsa}.
Typically, the number and size of Transformer layers is fewer than other domains, e.g., computer vision, which tend to be data-rich.
Interestingly, multi-layer perceptron (MLP) only modeling of human activities has seen renewed interest in recent months, through approaches such as MLP-HAR \cite{zhou2024mlp} and MLP-Mixer \cite{ojiako2023mlps}.

A defining feature of deep learning-based HAR is the promise of integrated feature learning, i.e., \textit{the features learned are specifically optimized for the task (HAR)}.  
The Feature Extraction and Classification steps in the canonical ARC (\autoref{fig:arc}) are combined into a single step.
This leads to HAR performance improvements over the aforementioned statistical and ECDF features. 
However, end-to-end training with deep learning-based methods requires substantial quantities of \textit{annotated data} for effective recognition, especially to utilize deeper networks.

% \vspace*{-.65em}
\subsection{Limits of Traditional Approaches including End-to-End Learning}
\label{sec:understanding}
Despite the substantial progress being made in (parts of) the modeling process, HAR is still a hard problem as evidenced by stagnating progress with regards to activity recognition accuracy in challenging scenarios.
Reasons for this plateau in progress can be summarized and categorized as follows.

\paragraph{Information coding} 
Recent, initial progress in CV was driven by deep CNNs which specifically exploit the way information is encoded in images: as a hierarchy of local geometric structures which not only represent signal level features but are also tightly connected to the semantics of different image components. 
By contrast, while multimodal sensor data can be represented as pseudo images with local structures, these structures are artifacts of the specific method used to create them from the sensor data. 
In general, they have little correspondence to the semantics of the data and the way information is encoded in the signal. 
As a consequence the impact of the deep CNN revolution on sensor based HAR has been limited. 
In particular the systems tend to be bad at generalising across data sets and users as even very similar sensor setups produce different fake image representations.

\paragraph{Lack of labeled training data} 
A key factor in the rapid improvement of CV methods has been the availability of nearly unlimited training data on various online platforms. 
While not all of this data is labeled, labeling images is relatively straight forward and can be  easily crowd sourced on a very large scale. 
As a consequence, data sets with millions of instances and thousands of classes are widely available. 
By contrast, while today huge amounts of multimodal sensor data are being produced by mobile, wearable and ubiquitous devices only a small fraction is openly available online. 
Furthermore, labeling sensor data is much more difficult than labeling images \cite{plotz2023if}. 
Anyone can distinguish a picture of a cat from a dog. 
Distinguishing on IMU signal for say squats and walking on the other hand can be difficult even for experts and is next to impossible for non experts. 
This means that crowd sourcing which has been so successful in CV is not on option for sensor data. 

As a consequence sensor based HAR labeled  data sets have orders of magnitude fewer classes (typically $\ll 100$) and instances (typically low double digits number of hours of data from just a few users). 
While much larger data sets with months of data from thousands of users have recently emerged, they are largely unlabeled. 

\paragraph{Signal ambivalence} 
Human models of the world are largely derived from visual perception. As a consequence in most cases visual information is needed to interpret situations, including activities in a way corresponding to human perception. 
Most sensors do not contain the same information

\section{The Importance of Representations}
\label{sec:ssl}
Even with the shift of the HAR community to end-to-end learning methods as outlined in the previous section, the field has still to witness those dramatic breakthroughs that other domains (CV, NLP, etc.) have seen.
Yet, the introduction of such DL methods that integrate feature learning components has underlined a central conclusion very strongly:
\textit{Representations of sensor data play a key role in the success (or lack of it) of HAR approaches.}

Initially, representations were learned mainly to overcome issues with heuristics that led to non-generalizable features \cite{plotz2011feature}.
Then the main focus shifted more towards compression as evidenced by the broader uptake of (variants of) auto-encoders (e.g., \cite{haresamudram2019role}). 
However, recently developed methods now focus on learning actual latent spaces that exhibit very promising capabilities with regard to the generalizability of the \textit{learned} representations -- with many new applications downstream \cite{haresamudram2022assessing}.

In the following main part of this tutorial, we will now survey contemporary techniques that specifically address the representation of movement sensor data that are the basis for HAR using wearables.

\subsection{Self-Supervised Representation Learning}

\begin{figure}
    \centering
    \includegraphics[width=0.65\linewidth]{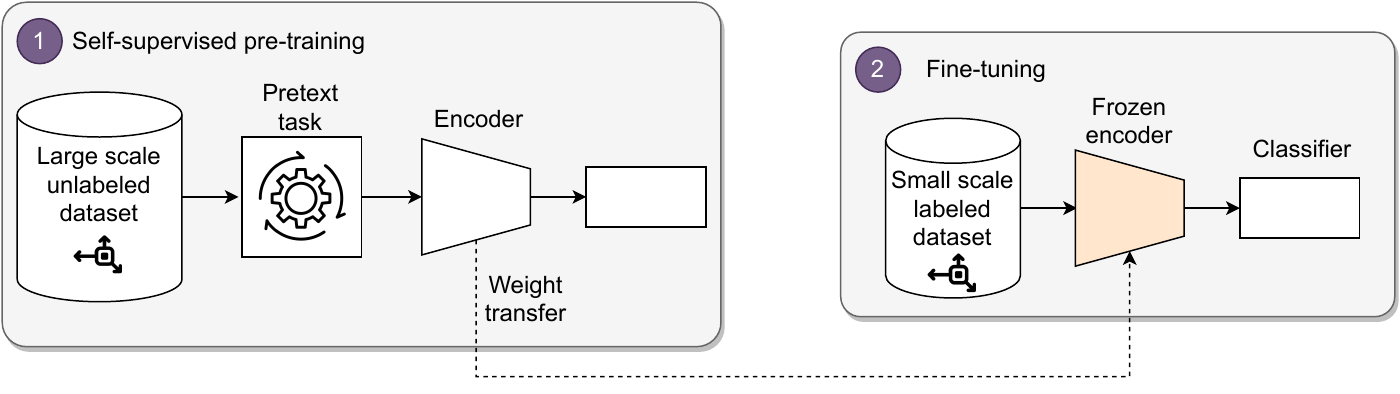}
    \caption{An overview of the self-supervised learning pipeline. 
    Reproduced with permission from \cite{haresamudram2022assessing}.
    }
    \label{fig:self_sup}
\end{figure}

For wearables applications, data collection is often performed through lab-based studies, especially if annotation needs to be performed. 
Participants are recruited to perform a handful of orchestrated activities (locomotion-style, e.g., walking, sitting, etc.) with on-body wearable sensor(s). 
Video is recorded synchronously \cite{plotz2012automatic} and used to annotate the streams of sensor data into activities \cite{roggen2010collecting, ciliberto2021opportunity++}.
This process is time consuming, expensive, and subject to privacy issues \cite{kwon2020imutube}.   
Due to these factors, publicly available wearable sensor datasets are typically limited in size and variability, often containing only 10-20 participants and a handful of activities, recorded over a few hours \cite{kwon2020imutube}.
This hinders the development of complex and truly deep neural networks.

However, simply collecting large quantities of \textit{unlabeled} wearable sensor data is straightforward. 
Smartwatches can be shipped out to thousands of participants for a few days to perform data collection, resulting in truly in-the-wild data, albeit without much control over data quality and without knowledge about the activities being performed. 
This protocol was, for example, utilized in the UK Biobank study, which recorded accelerometer data from approx.\ $90k$ participants, resulting in around 20TB of sensor data \cite{doherty2017large, willetts2018statistical}. 
As a result of this \textit{more diverse} data collection, the participant pool is not limited geographically to people that are close to laboratories where collection is performed.
Furthermore, wearable devices and sensors evolve over time -- their underlying architecture and designs change, they become more compact, consume less power, and become more powerful.
Accordingly, the sensor data distributions change over time, and it becomes prohibitively expensive to initiate data annotation efforts for each new sensor, as previous models can fail due to distribution shifts.
Overall, learning from unlabeled data can be more advantageous for wearables, as it is possible to not only learn from larger scale data but also from more diverse data, without requiring any annotation \cite{gidarissCVPR2021}.

The wearables community has therefore moved on from studying deeper supervised neural networks, towards opportunistically leveraging available and more easily collected unlabeled data for activity recognition.
This paradigm is called \textit{Self-Supervised Learning}, where (large-scale) unlabeled data are first utilized to learn useful, generic representations (i.e., neural network weights).
Subsequently, the learned weights are optimized to the actual downstream task (e.g., recognizing daily activities), using much smaller-scale annotated data. 
Therefore, once pre-training is complete, the learned weights can be repeatedly fine-tuned/used to extract features for numerous downstream applications. 

This training paradigm is described as `pretrain-then-finetune' \cite{haresamudram2022assessing}, and comprises two steps (as shown in Fig.\ \ref{fig:self_sup}): \emph{(i)} pre-training -- where the network is trained to solve a different but (hopefully) useful \textit{pretext} task using only unlabeled data.
This task requires some \textit{semantic understanding of the data/domain to solve,} thereby resulting in useful representations; and 
\emph{(ii)} fine-tuning/classification -- where the pre-trained weights are either used directly for feature extraction for HAR, or they are further fine-tuned to recognize the activities under study.

The design of suitable pretext tasks is vital for learning useful representations.
The task cannot be too easy, lest the network learns nothing useful by solving it; similarly, the task cannot be excessively difficult to solve, making it too challenging to learn useful representations \cite{haresamudram2021contrastive}.
For example, Contrastive Predictive Coding (CPC) \cite{oord2018representation, haresamudram2021contrastive} utilizes a contrastive future prediction task, and it was found that predicting one or two timesteps in the future is (too) easy and results in poor performance, whereas predicting multiple timesteps into the future is much more complex, and leads to very effective representations. 
A number of such tasks have been developed, typically withholding parts of the input from the network, and training the network to predict the missing data \cite{gidarissCVPR2021}. 
For example, masked autoencoders \cite{he2022masked, huang2022masked} set portions of the data to zero and the task is to reconstruct the missing portion, based on available context.
In what follows, we detail and discuss three early self-supervised methods in the community, which laid a foundation for applying self-supervision to wearables applications.

\subsubsection{Autoencoders}
\iffalse
\begin{figure}
    \centering
    \includegraphics[width=0.8\linewidth]{Figures/autoencoder.pdf}
    \caption{
    Caption.
    \harish{keep this figure and make new ones for the rest? Or not?}
    }
    \label{fig:enter-label}
\end{figure}
\fi

These models represent one of the earliest unsupervised methods employed for sensor-based HAR.
Here the pretext task involves reconstructing windows of sensor data, after passing through a series of network layers with varying sizes. 
Typically, there are two major components in the architecture -- \textit{(i)} the Encoder; and \textit{(ii)} the Decoder.

The encoder is used to encode the input sensor data into embeddings by passing input data through a cascade of layers with ever reducing size--forming an information ``bottle neck''--down to a lower-dimensional, compact internal representation.
The decoder, which is usually a mirror image of the encoder, then reconstructs the input data from the internal embeddings through step-wise dimensionality increase up to the original dimension. 
A variety of layers have been used in autoencoders, including fully connected \cite{haresamudram2019role}, convolutional (1D and 2D), and recurrent \cite{haresamudram2019role, abedin2019sparsesense} layers.
Reconstructing the input after passing through smaller layers creates aforementioned information bottleneck, forcing the network to learn salient features \cite{goodfellow2016book}. 
Given a window of sensor data $W$, an encoder $g$ and a decoder $d$, the loss function is defined as: 
$$ L(W; g, h) = ||W - d(g(W))] ||^2_{Fro}$$

\subsubsection{Multi-Task Self-Supervised Learning}
This is the first self-supervised approach introduced for wearables-based HAR. 
It relies on a set of eight data transformations / augmentations introduced by Um \etal \cite{um2017data} for increasing sensor data diversity, including adding random Gaussian noise, scaling, rotations, negation, flipping channels, permuting sub-segments of windows, time-warping and channel shuffling. 
For windows of sensor data, Saeed \etal \cite{saeed2019multi} applied each transformation with a probability of 50\%, and passed the probabilistically transformed windows through a common convolutional encoder.
Subsequently, MLPs are applied separately to each branch, and used for classifying whether each branch has had the transformation applied or not. 
The final loss is the sum of the losses from each branch, and used for updating model weights. 
The intuition behind this pretext task is that it captures core signal characteristics and sensor behavior under different rotations and placements, and levels of noise \cite{saeed2019multi}.
This results in strong performance, relative to newer methods, as shown by Haresamudram \etal \cite{haresamudram2022assessing}.
Tang \etal \cite{tang2021selfhar} expanded this self-supervised objective with self-training, assuming semi-supervised learning scenarios when there is a large amount of unlabeled data with limited labeled ones at training time. The use of a teacher-student knowledge distillation setup and confidence score filtering enables an efficient semi-supervised learning framework in which only high-quality unlabeled data are used for training, and improved performance was shown compared to single-dataset training scenarios with the use of publicly-available datasets. However, the authors noted that bias in the teacher model can impact performance in extremely low data-availability scenarios. 

\subsubsection{Masked Reconstruction}
An extension of the autoencoder setup involves masking portions of the input data, and training the network to reconstruct only the missing portion from available context, e.g., context encoders \cite{pathak2016context} and BERT \cite{devlin2018bert}.
For wearables, masked reconstruction \cite{haresamudram2020masked} utilizes transformer encoder \cite{vaswani2017attention} layers and masks out 10\% of the sensor data of an analysis windows at random timesteps.
As such, a window of data $W$ is perturbed using a binary mask $M$, and then passed through the transformer encoder layers $g$.
A set of fully connected layers $h$ is used to match the dimension to the input, and the mean squared error (MSE) loss is computed only on the masked portion of the sensor windows. 
The loss is defined as: 
$$ L(W, M; g, h) = || (1-M) \odot [W - h(g(M \odot W))] ||^2_{Fro}$$ where $\odot$ denotes element wise multiplication.

The masking creates a mismatch between training and testing conditions (where there is no perturbation).
For each of the chosen timesteps, processing is as follows: 
\textit{(i)} with a probability of 80\%, the data are set to zero (i.e., the masking is performed); 
\textit{(ii)} the data are left unchanged with probability of 10\%; 
and 
\textit{(iii)} the data are replaced with a random timestep from within the frame with probability of 10\%. 
This strategy is useful for reducing the impact of differing training and testing conditions. 
Overall, the masked reconstruction setup can be interpreted as a denoising autoencoder, where the input data are intentionally noised first, and the network is trained to reconstruct clean data instead.
The number of timesteps of sensor windows to be masked is a hyperparameter, where tuning leads to improved performance \cite{haresamudram2022assessing}.
In addition, masking across both sensor channels and timesteps is also an effective option \cite{miao2024spatial}.
Alternatively, Hong \etal \cite{hong2024crosshar}, 
improve cross dataset performance via masked modeling in conjunction with contrastive regularization.

\subsection{Representation Alignment and Structuring}
\subsubsection{Contrastive Learning and Siamese Networks for Human Activity Recognition}
\label{sec:contrastive}

\begin{figure}
    \centering
    \includegraphics[width=0.8\columnwidth]{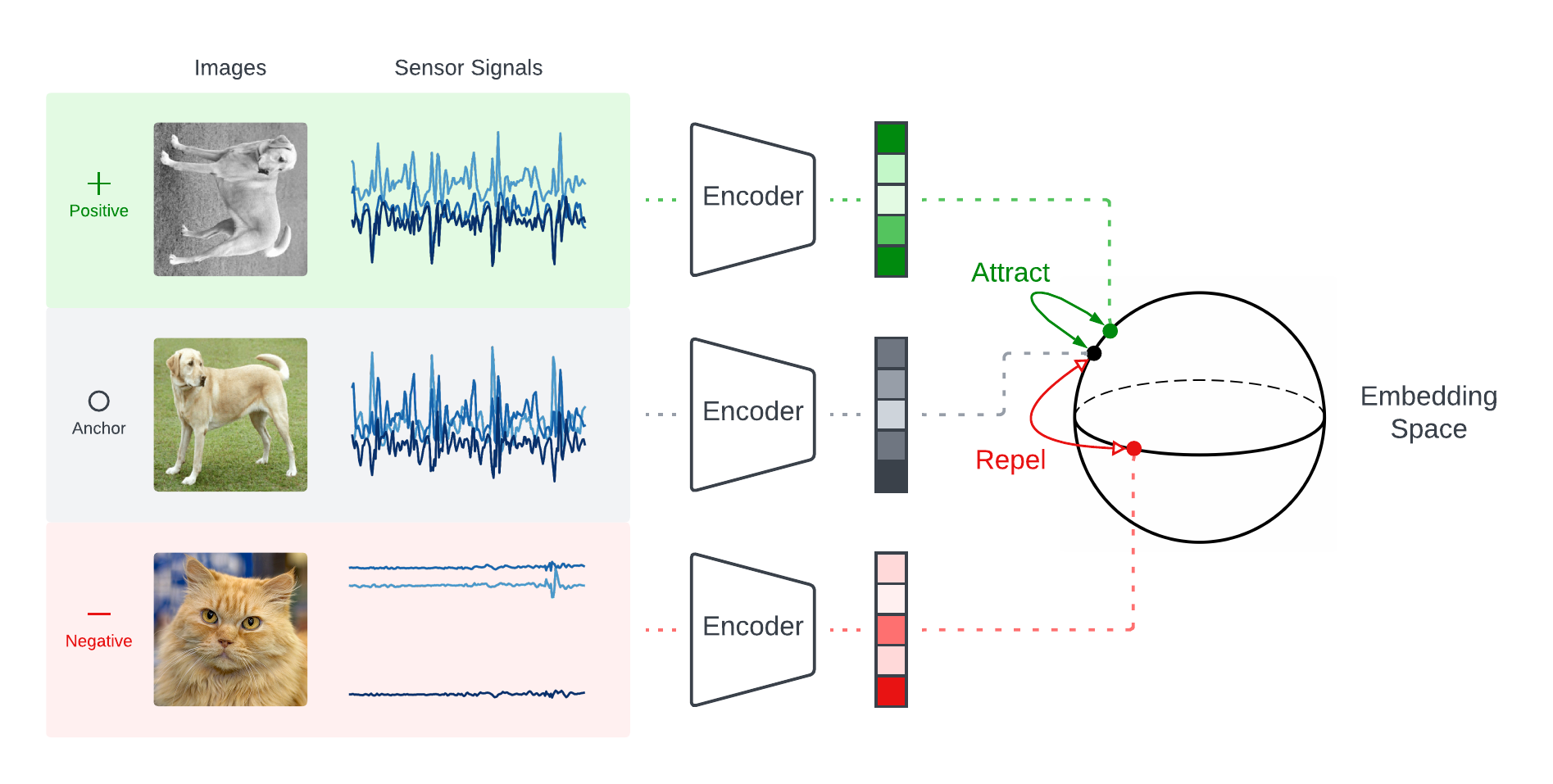}
    \caption{An overview of contrastive learning (CL). Data samples (images or sensor signals depending on the target task) are first passed through an encoder to obtain embeddings in a latent space. These embeddings are pushed closer to each other or pulled apart depending on whether the embedding forms a positive pair or a negative pair with the anchor.}
    \label{fig:contrast_general}
    \vspace*{-1em}
\end{figure}

Siamese networks are based on the concept that twin sub-networks connected at the output layer should generate similar outputs for a pair of similar but distinct inputs.
They have been proposed in the early 1990s, primarily for tasks such as signature verification~\cite{becker1992self, bromley1993signature}. 
These sub-networks usually share the same set of weights and utilize a distance function at the output as the loss function, aiming to reduce the distance between embeddings of similar pairs. 

Building on the foundation of Siamese networks, contrastive learning (CL) emerged as a robust training paradigm in self-supervised learning. 
This approach extends the Siamese architecture by also defining negative pairs, where the model is trained to increase the distance between embeddings (Fig.\ \ref{fig:contrast_general}). 
Many studies and training frameworks have demonstrated CL's efficacy in developing powerful feature extractors without the need for labels \cite{oord2018representation, linear_bachman2019learning, chen2020simple, chen2020improved}. 
CL is based on the ability to identify data clusters with high intra-group similarity and low inter-group similarity, leveraging these characteristics to train models that effectively distinguish between similar and dissimilar samples in the embedding space.

As an illustrative example, consider a scenario where the model needs to distinguish between images of dogs and cats (see Fig.\ \ref{fig:contrast_general} with images as input). 
In traditional supervised learning, labeled images of dogs and cats are fed to the model with their corresponding labels. 
In contrastive learning, pairs of images are presented to the model, in which the model is trained to generate similar embeddings if they belong to the same category (positive pair) or generate dissimilar embeddings if they are from different categories (negative pair). 
In this process, the method used to generate positive and negative pairs is critical, since it encodes semantic meaning that we want the model to capture. 
Chen \etal \cite{chen2020simple} proposed SimCLR, a simple contrastive learning framework that leverages augmentations to encode such semantic meaning. 
In particular, two images form a positive pair if one is a transformed version of the other, and form a negative pair if they are not originally the same image. 
Despite its simplicity, it outperforms other self-supervised learning techniques on commonly used CV benchmarks, demonstrating that the model learns rich, discriminative features without relying on explicit labels.

\subsubsection{Adaptations of Contrastive Learning in Human Activity Recognition}
\label{sec:contrastive_adaptations}

\begin{figure}
    \centering
    \includegraphics[width=0.8\columnwidth]{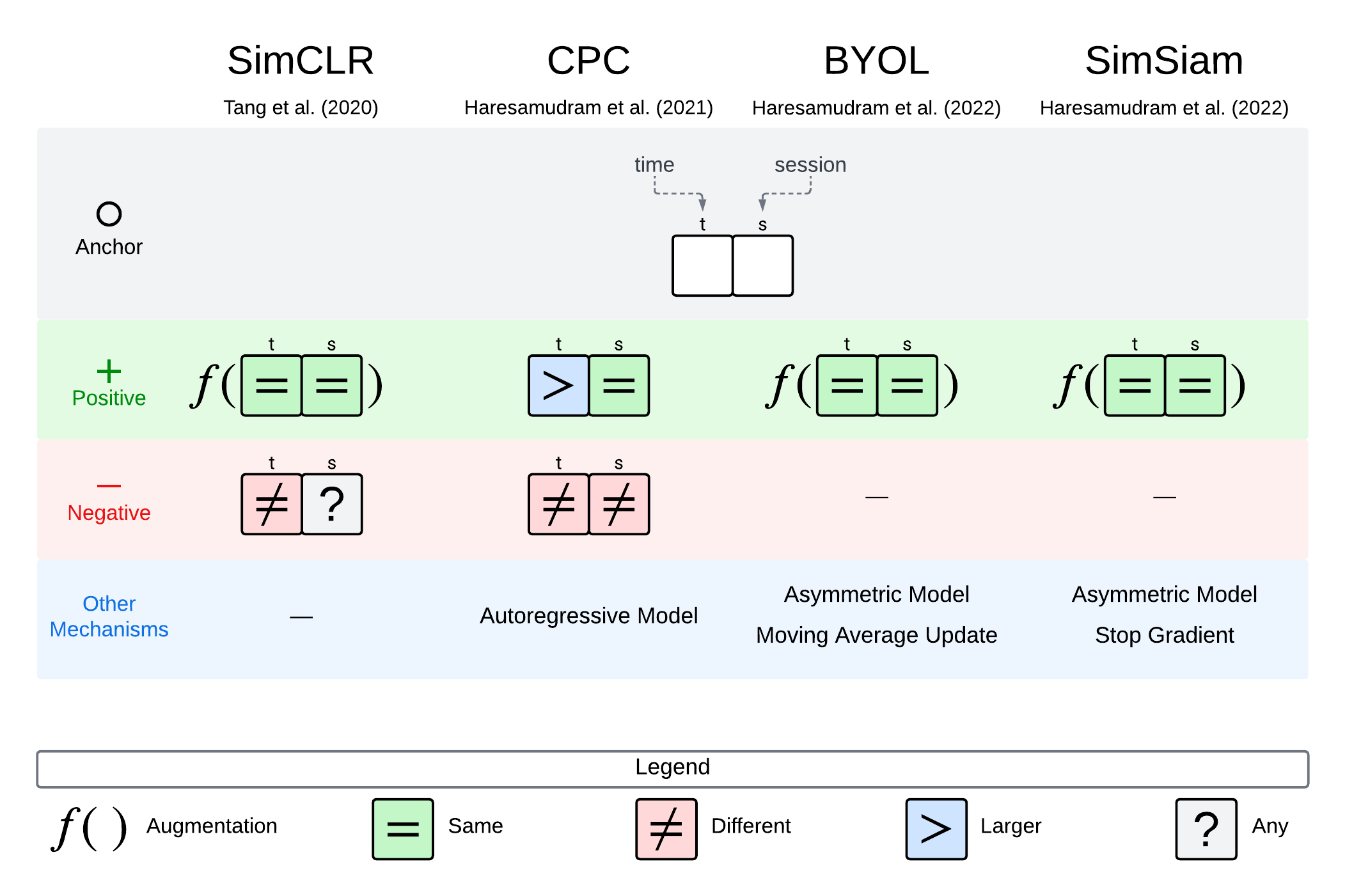}
    \caption{Contrastive learning adaptations in human activity recognition. A comparison is drawn among SimCLR~\cite{tang2020exploring}, CPC~\cite{haresamudram2021contrastive}, BYOL~\cite{haresamudram2022assessing} and SimSiam~\cite{haresamudram2022assessing}. A high degree of commonality can be found among these frameworks, especially in the use of augmented anchors as positive samples for SimCLR, BYOL, and SimSiam. The CPC differs from the rest by using future samples instead of augmented views. SimCLR and CPC leverage time-misaligned samples as negatives, while BYOL and SimSiam leverage additional mechanisms to remove the requirement of negative samples.}
    \label{fig:contrast_single}
    \vspace*{-1em}
\end{figure}

Even though many of the aforementioned methods have been developed in, for example, the computer vision research community, they are, strictly speaking, modality-agnostic:
More generally, contrastive learning can be viewed as learning on the augmentation graph on data samples \cite{haochen2021provable}. 
As a result, there have been various efforts to adapt and extend contrastive learning techniques to the area of human activity recognition. 
An overview of such methods is given in Fig.\ \ref{fig:contrast_single}.

\paragraph{SimCLR} One such adaptation effort is the modification of the SimCLR framework for sensor time series by Tang \etal \cite{tang2020exploring}. 
While the overall framework remains identical, instead of images, sequences of sensor data from wearable devices become the input (Fig.\ \ref{fig:contrast_general} and \ref{fig:contrast_single}). 
The model is now tasked with distinguishing between positive and negative activity samples that are generated using augmentation functions. 
Challenges in such adaptation remain in incorporating the uniqueness of sensor signals and the temporal and sequential nature of the data. 

Since the choice of augmentation function encodes the learning objective and invariance that the model should capture, the original SimCLR framework developed for computer vision tasks demonstrates significant performance differences among the choice of augmentation functions \cite{chen2020simple}. 
Unlike images, sensor data captures dynamic patterns over time, and therefore image transformation functions would not be suitable for time series. 
Temporal dependencies, varying speeds of movements, and nuanced transitions between activities demand a specialized approach. 
In the absence of labeled data, the model must discern subtle differences in motion patterns and learn robust representations of human movements. 
In the adaptation \cite{tang2020exploring}, the authors explored different combinations of commonly used augmentation functions for time series, including adding Gaussian noise, applying a random 3D rotation, time-warping, etc. 
The authors demonstrated that the choice of augmentation can have a significant impact on model performance in HAR, with the 3D-rotation augmentation demonstrating the best performance when the model is fine-tuned. 
This work showcased improvements over supervised and unsupervised learning methods, highlighting the potential benefits of contrastive learning in HAR systems, such as improved clustering of activities \cite{ahmed2022clustering}.
However, further exploration is warranted to determine the conditions under which such adaptation proves advantageous for HAR and other healthcare-related applications.

\paragraph{Contrastive Predictive Coding} 
In parallel to SimCLR, the endeavor to leverage unlabeled sensor data for HAR has sparked interest in other contrastive learning techniques, notably Contrastive Predictive Coding (CPC). 
Haresamudram \etal \cite{haresamudram2021contrastive} adapted CPC to HAR, emphasizing the importance of temporal structure in sensor data representation. 
The original formulation for CPC \cite{oord2018representation} is motivated by the predictive coding theory in neuroscience in which the brain constantly generates hypotheses and updates the mental model of the environment by comparing the hypotheses with the sensory inputs. 

In the CPC framework, the authors modeled this as a machine learning task. 
The model is separated into three main components: the data encoder, which encodes input data into latent representations, the autoregressive network, which summarizes past latent vectors into a single context vector, and the prediction network, which predicts future samples based on the context vector. 
A probabilistic contrastive loss is used to provide a supervisory signal for the learning process, in which the model is optimized to maximize the mutual information between the future sample and the context vector. 
As a result, by formulating CPC as a contrastive prediction task, the model learns effective representations by exploiting the inherent dependence between data samples. 

The adaptation by Haresamudram \etal \cite{haresamudram2021contrastive} focuses on leveraging the temporal dependence that is present in sensor time series, in which it is possible to naturally formulate a predictive task for a continuous stream of sensor data. 
Their work demonstrates the practical value of CPC-based pre-training, showcasing improved recognition performance even with small amounts of labeled data. 
Furthermore, investigations into enhancements of CPC for HAR, as presented by subsequent research \cite{haresamudram2023investigating, haresamudram2024towards}, have yielded fully convolutional architectures that exhibit substantial improvements across diverse datasets and activity types. 
These advancements underscore CPC's potential to empower a wide range of HAR applications, with theory-backed training frameworks.

\paragraph{BYOL and SimSiam} 
Beyond these early efforts in adapting CL to HAR, Haresamudram \etal \cite{haresamudram2022assessing} conducted a comprehensive comparison of different contrastive and Siamese learning methods for HAR. 
These included other contrastive and Siamese approaches such as BYOL \cite{grill2020bootstrap} and SimSiam~\cite{chen2021exploring} (also summarized in Fig.\ \ref{fig:contrast_single}). 

BYOL, which stands for Bootstrap Your Own Latent, was proposed as an alternative to contrastive learning to remove the requirement for negative samples and large batch sizes. 
Without negative samples, model collapse can happen, where the model produces a trivial representation for all inputs. 
This is avoided by the use of asymmetric model architectures and exponential moving average updates in the BYOL framework. 
In a follow-up work, SimSiam \cite{chen2021exploring} aims to identify a minimal setup for Siamese learning that can deal with model collapse. 
The authors demonstrated that the combination of asymmetric model architectures and the stop-gradient operation can prevent model collapse, and the mean-squared error works well as a loss function. 

Although these frameworks demonstrated superior performance compared to contrastive approaches in computer vision tasks, Haresamudram \etal \cite{haresamudram2022assessing} showed that the SimCLR framework \cite{tang2020exploring} and multi-task self-supervised learning \cite{saeed2019multi}, outperformed other approaches, including BYOL, SimSiam, masked reconstruction, Contrastive Predictive Coding (CPC) and autoencoders. 
This indicates that the effectiveness of different techniques can differ between different modalities. 
In addition to comparing different training frameworks, they also explored model performance across other dimensions, providing insights into the robustness, dataset characteristics, and feature space characteristics of self-supervised methods. 
By evaluating seven state-of-the-art self-supervised techniques for HAR, this study contributes to understanding the value of self-supervised learning in learning representations for diverse scenarios in HAR.
Qian \etal \cite{qian2022makes} also study contrastive training for small scale wearable datasets, with the goal of discovering key components to learn more effective representations. 
The pre-training data efficiency of self-supervision was evaluated in \cite{dhekane2023much}, which found that even a few minutes of unlabeled data with sufficient augmentation can rival using entire dataset for pre-training.

\subsubsection{Multi-device and Multi-modal Contrastive Learning for Human Activity Recognition}
\begin{figure}
    \centering
    \includegraphics[width=0.8\columnwidth]{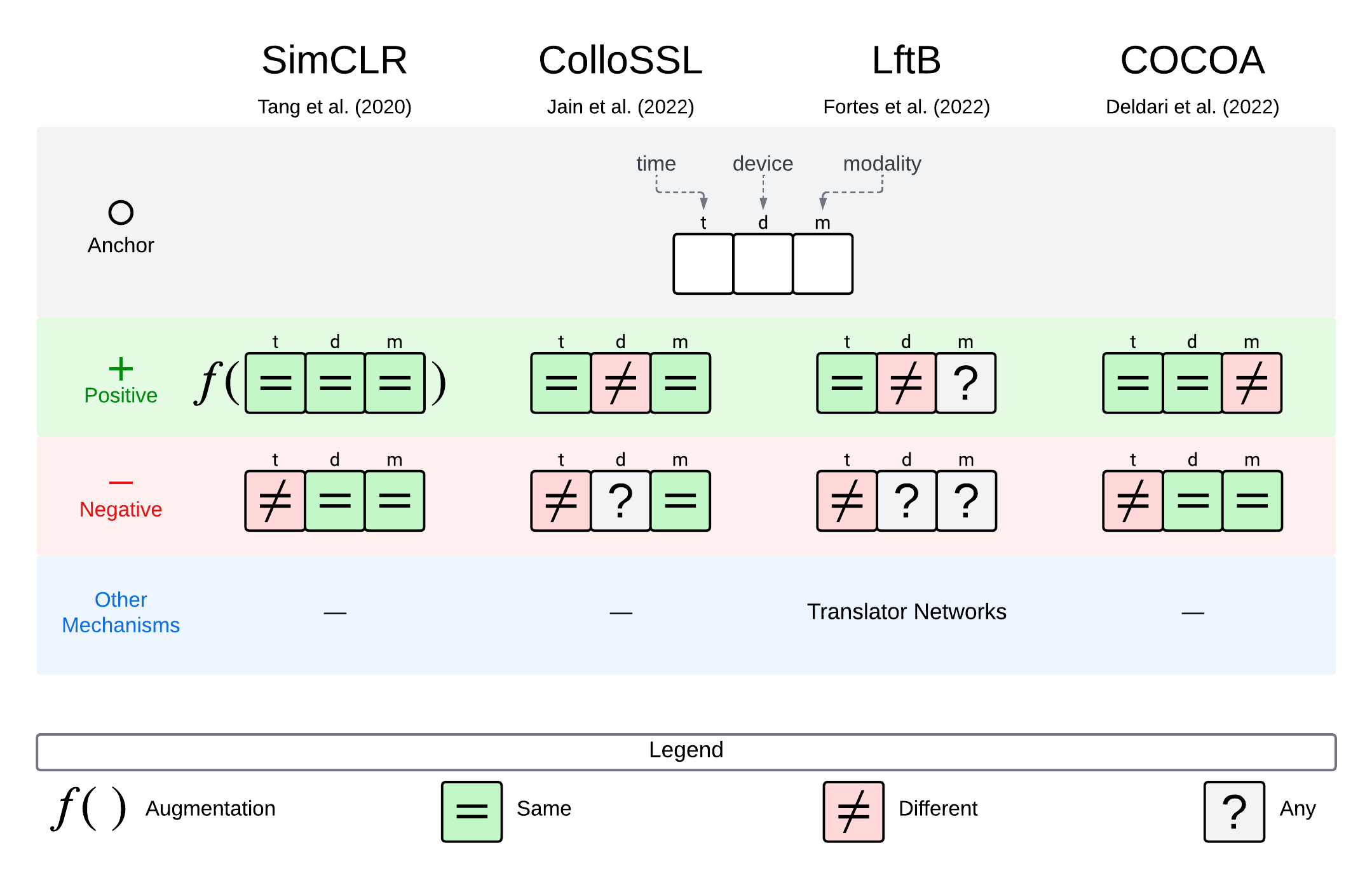}
    \caption{An overview of multi-device and multi-modal contrastive learning frameworks in human activity recognition. A comparison is drawn among ColloSSL~\cite{jain2022collossl}, Learning from the Best (LftB)~\cite{fortes2022learning}, COCOA~\cite{deldari2022cocoa} and SimCLR~\cite{tang2020exploring} (as a single-device single-modality reference). Instead of relying on augmentations (as in SimCLR), these multi-device and multi-modal approaches leverage data from time-aligned data from different devices (ColloSSL and LftB), and different modalities (COCOA) for positive samples. Time-misaligned samples as negatives is a common feature among these approaches, with different frameworks imposing additional limitations on the sampling algorithm.}
    \label{fig:contrast_multi}
    \vspace*{-1em}
\end{figure}

In the previous sections, we have surveyed examples of adaptations of contrastive learning approaches in human activity recognition. 
Even though these contrastive and Siamese frameworks are modality-agnostic and shown to be effective, they do not leverage the modality-specific characteristics of sensor time series.

In the more recent works, the community proposed frameworks that leverage the naturally occurring transformations and synchronization inherent in multi-device and multi-modal sensor data for contrastive learning (summarized in Fig.\ \ref{fig:contrast_multi}). 
Instead of using hand-picked, artificial augmentations like that like SimCLR and BYOL \cite{tang2020exploring, haresamudram2022assessing, chen2020simple, grill2020bootstrap}, the ColloSSL (Collaborative Self-Supervised Learning) framework \cite{jain2022collossl} introduces a novel method to tackle the scarcity of labeled data in Human Activity Recognition by leveraging time-synchronized data collected from multiple inertial sensors worn by users. 
The key insight lies in the observation that different devices worn by the same user capture the same physical activity -- just from different perspectives, and are natural transformations of each other. 
This allows an intuitive formulation of contrastive learning in which positive samples come from time-aligned samples from different devices, while negative samples come from time-misaligned samples. 
ColloSSL employs a combination of device selection and contrastive sampling algorithms to form training batches from multiple devices, enabling contrastive learning without labeled data. 
A multi-view contrastive loss function, which extends traditional contrastive learning to a multi-device setting, was also used in this approach. 
It was demonstrated to have superior performance in standard evaluation setups and low-data regimes compared to conventional supervised and semi-supervised methods. 
This indicates the potential for leveraging multi-device setups for more effective training.

Also leveraging time-synchronized sensor time series but with a different objective, `Learning from the Best` (LftB) \cite{fortes2022learning} adopts a flexible approach to cross-device contrastive learning by using individual encoder and translator networks for each sensor location to separate the embedding spaces from devices. 
To achieve the goal of improving the quality of the feature extractor for a target device, this scheme leverages a cross-domain contrastive learning setup from domains that are only available during training. 
In particular, data are first passed through their corresponding encoders, and then through pairwise translator networks, which translate embeddings from one device to another. 
The InfoNCE loss \cite{oord2018representation} is used again for CL, where the representation from the target domain is contrasted against another that was translated from a different domain.
The authors further proposed to reuse the translator networks for classifier training, in which the data from other domains are translated to train the classifier, in addition to using data from the target domain. 
Evaluated on the HAR benchmark datasets, this method shows improvements in F1 scores, particularly for activities that benefit most from additional sensor information. 
By leveraging contrastive learning to enhance target sensor performance, it addresses the limitation of relying solely on one single device that might be subject to more motion artifacts.

Another work, COCOA (Cross Modality Contrastive Learning) \cite{deldari2022cocoa}, looks at the multi-modality aspect of sensor time series. 
It introduces a contrastive learning method that leverages synchronous data segments from different sensor modalities to create positive samples, instead of relying on augmented or temporally related pairs. 
Similar to ColloSSL, this method takes advantage of the synchronous nature of multi-modal data to enhance the learning process and it incorporates dynamic sensor selection to ensure the quality of the positive and negative pairs. 
Due to the differences in signal patterns among data modalities, the authors proposed the use of modality-specific encoders, which improve the model's robustness to missing modalities and its computational efficiency. 
This approach uses a different formulation of the contrastive loss: a two-part loss function that is dedicated to maximizing inter-modality agreement while minimizing the agreement between temporally distant samples. 
Evaluations show that COCOA outperforms state-of-the-art models in various tasks, such as human activity recognition, sleep stage detection, and emotion recognition, especially when there are more than two data modalities, demonstrating its efficacy and generalizability across different types of sensor data.
Finally, some works also explore the possibility of contrastive training between wearable movement sensors and pose data \cite{choi2023multimodal}.

These advancements in leveraging multi-device and multi-modal sensor data represent a community effort in driving forward the field of HAR. 
By moving away from traditional, artificially augmented data and instead utilizing the natural, synchronous relationships inherent in multi-sensor setups, these methods extract more meaningful and robust features. 
This allows us to better utilize the modality-specific characteristics of sensor time series. 
The positive results across HAR tasks also indicate that these approaches could be refined and expanded to other domains, thereby enhancing the applicability and impact of contrastive learning in real-world scenarios.

\subsubsection{Contrastive Learning outside Human Activity Recognition}
In addition to HAR, CL was successfully applied in other mobile sensing tasks, such as change point detection \cite{deldari2021time} and emotion recognition \cite{dissanayake2022sigrep}.

One such work looked at the use of contrastive predictive coding (CPC) to detect changes in web service traffic and mobile application usage, in addition to human activity recognition \cite{deldari2021time}. 
The authors proposed $TS-CP^2$ (Time Series Change Point detection method based on Contrastive Predictive coding), which leverages CL to detect changes in time series by using time-adjacent intervals as positive pairs and those separated across time as negative pairs. 
This method enhances change point detection by utilizing contrastive learning to capture the inherent temporal dependencies within time series data. 
Experiments on datasets have shown that $TS-CP^2$ outperforms other change point detection methods, highlighting the effectiveness of contrastive learning in capturing subtle changes in time series properties without relying on labeled data. 

SigRep \cite{dissanayake2022sigrep} studied emotion recognition from wearable physiological signals using contrastive learning. 
This method focuses on learning robust representations of different data modalities, such as heart rate and electrodermal activity, captured from consumer-grade wearable devices. 
By leveraging contrastive learning, SigRep contrasts augmented signal samples to create robust feature representations for different modalities, which can be used for downstream emotion classification tasks. 
Evaluated on publicly available datasets, the method demonstrated superior performance in emotion recognition compared to state-of-the-art methods. 
Additionally, it showed resilience to signal losses and required fewer labeled data for effective training.

\subsubsection{Adversarial Learning Based Representation Alignment}

 \begin{figure}[!t]
    \centering
    \includegraphics[width=0.7\columnwidth]{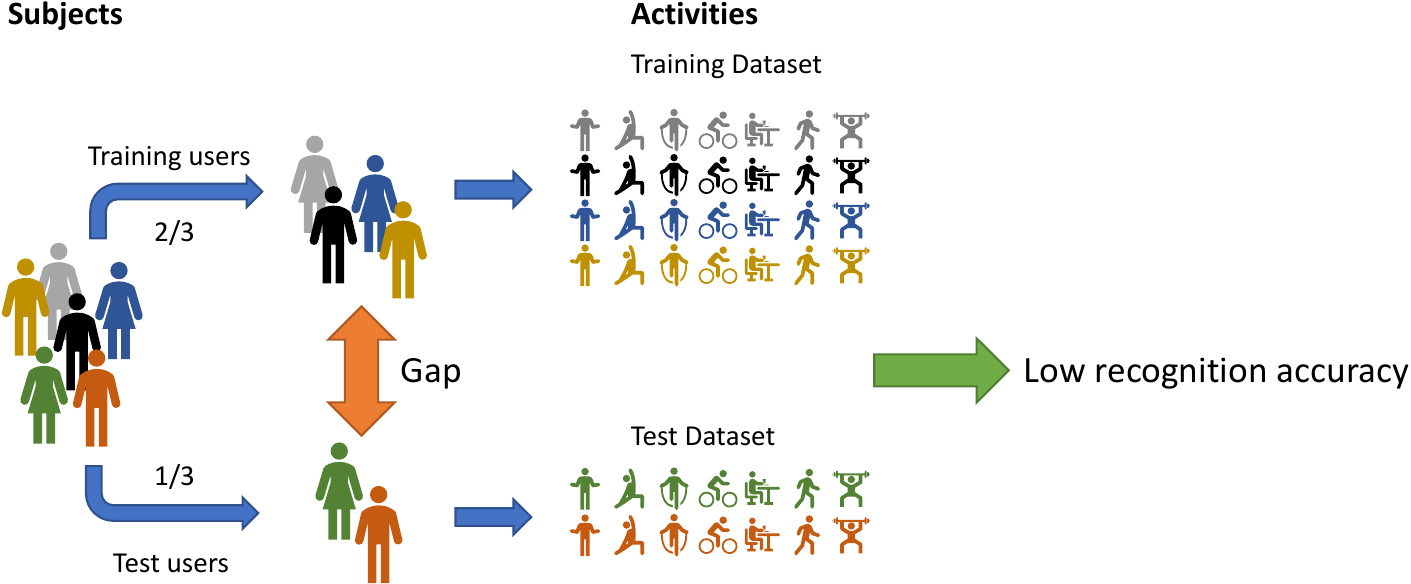}
    \caption{Challenges in activity recognition: accounting for diverse behavior patterns across individuals.}
    \label{fig:adversarialmotivation}
    \vspace*{-1em}
\end{figure}

HAR faces challenges in handling data from diverse subjects and generalizing to unseen users. 
Studies \cite{soleimani2021cross, suh2022adversarial, kang2022augmented, suh2023tasked} highlight the impact of individual characteristics and behaviors on HAR performance, with significant degradation when models are applied to new users (\cref{fig:adversarialmotivation}).

Two primary categories of approaches address this issue: classic, and deep learning-based methods. 
The former involve selecting user-invariant features or building user-specific models, which--while effective--pose challenges in terms of labeled data availability and potential performance trade-offs. 
In response, DL techniques, particularly multi-task and generative adversarial learning (GAN), have been used to tackle data distribution challenges.

Chen \etal \cite{chen2020metier} introduced the METIER model, employing deep multi-task learning for activity and user recognition. 
By sharing parameters between activity and user recognition modules, they demonstrated improved performance through a mutual attention mechanism in the user recognition module. 
While effective, the generalization capability of these models beyond the training subjects remains unclear. 
Sheng \etal \cite{sheng2020weakly} proposed weakly supervised multi-task representation learning using Siamese networks, mitigating environmental differences through similarity-based multi-task learning. 
However, their representation tends to create subject-specific clusters, potentially hindering generalization. 
In contrast, Bai \etal \cite{bai2020adversarial} leveraged adversarial learning to generate robust feature representations regarding user variations. 
Using Wasserstein GAN and Siamese networks, they demonstrated the ability to generalize to new subjects without sacrificing performance, addressing concerns about neural network information leakage.

Despite these advancements, limitations persist. 
Adversarial learning, while enhancing performance, lacks a mechanism to measure the degree of generalization during training. 
Furthermore, such methods \cite{bai2020adversarial, leite2020improving, soleimani2021cross} may suffer when the feature extractor or generator overly focuses on fooling the discriminator.

To address these challenges, Suh \etal \cite{suh2022adversarial, suh2023tasked} proposed a cross-subject adversarial learning approach for sensor-based HAR.
This model learns subject-independent embeddings through adversarial learning, capable of generalizing to new subjects. 
The Maximum Mean Discrepancy (MMD) regularization quantifies feature generalization, enhancing the model's ability to generalize across subjects. 
The architecture \cite{suh2022adversarial} utilizes an encoder-decoder structure based on CNN, preserving signal characteristics. 
TASKED \cite{suh2023tasked} extended this with a transformer network \cite{dosovitskiy2020image, plizzari2021skeleton}, accounting for sensor orientations and spatial-temporal features. 
The inclusion of teacher-free self-knowledge distillation \cite{yuan2020revisiting} improves training stability, balancing feature generalization, and activity recognition optimization. 
In this method, self-knowledge distillation not only prevents overfitting but also guards against bias in cross-subject feature generalization induced by adversarial learning and MMD regularization. 
This comprehensive approach represents a significant stride in overcoming limitations and advancing the state-of-the-art in sensor-based HAR.

\section{Data Generation and Augmentation}
\label{sec:generation}

\subsection{Data Augmentation for HAR}

\begin{figure}[!t]
    \centering
    \includegraphics[width=0.6\columnwidth]{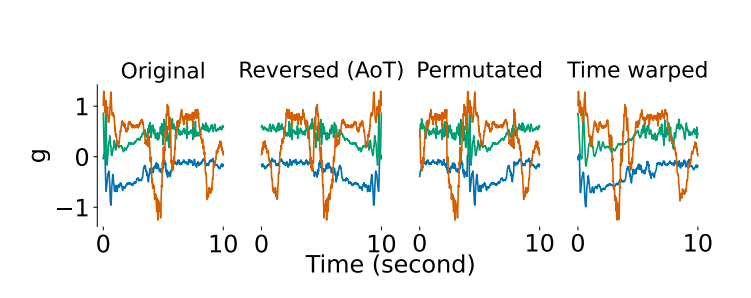}
    \vspace*{-1em}
    \caption{Examples of data augmentation for time-series data.}
    \label{fig:dataaugmentation}
    \vspace*{-1em}
\end{figure}

The effectiveness of DL models often depends on the availability of extensive datasets for training purposes. 
However, acquiring high-quality training data for HAR, particularly from wearable sensors, presents unique challenges. 
Unlike computer vision and sound classification, where ground-truth data can be readily obtained from online platforms like Amazon's Mechanical Turk, HAR data necessitates direct offline collection from users' physical behaviors. 
This process is time-consuming and labor-intensive, leading to a scarcity of labeled training data due to the inadequacy of publicly available datasets.

Recognizing the crucial role of data quantity in model performance, researchers have utilized data augmentation techniques, which are a popular technique to address the limitations imposed by insufficient training data \cite{kim2021label, jeong2021sensor}. 
While traditional CV approaches employ simple affine transformations for data augmentation, such as translation, rotation, resizing, and shearing \cite{shorten2019survey, maharana2022review}, the unique characteristics of accelerometer signals in sensor-based HAR necessitate alternative strategies. 
Additionally, image synthesis methods that blend foreground objects, background images, etc., are used to generate training examples \cite{dwibedi2017cut, yun2019cutmix}. 
Signal-processing methods (e.g., time-stretching, pitch-shifting, and dynamic-range compression) are applied to augment audio signals in sound classification tasks \cite{cui2015data, salamon2017deep, nanni2020data}.

In sensor-based HAR, distinguishing between foreground and background categories in accelerometer signals poses a challenge, rendering conventional image synthesis methods impractical. 
However, leveraging the temporal nature of accelerometer signals allows for the application of augmentation methods grounded in signal processing \cite{um2017data, ohashi2017augmenting, mathur2018using, steven2018feature, chung2019sensor, kalouris2019improving, cheng2023learning}. 
These methods include jittering, scaling, rotation, and random sampling, which have demonstrated efficacy in enhancing the diversity of the training dataset.

To address the limitations posed by the physical constraints of body-worn sensors, Ohashi \etal \cite{ohashi2017augmenting} introduced a data augmentation method tailored to the two-axis rotation capabilities of armband sensors. 
This approach outperformed conventional methods, emphasizing the importance of considering sensor-specific constraints in augmentation techniques. 

Furthermore, researchers have explored augmentation strategies to address specific challenges in data collection. 
Um \etal \cite{um2017data} tackled the scarcity of labeled motor states in Parkinson's disease patients by introducing a variety of augmentation techniques, including jittering, scaling, rotation, and permutation (\cref{fig:dataaugmentation}). 
Their findings highlighted the effectiveness of rotations, particularly in capturing the variability of sensor placement, which significantly improved the performance of Convolutional Neural Network (CNN) models.
Considering software and hardware heterogeneity affecting sensor data, Mathur \etal \cite{mathur2018using} introduced a data augmentation method that accounted for timestamp jittering caused by variations in accelerometer signal sampling rates. 

In the context of senior adults' physical activity recognition, Kalouris \etal \cite{kalouris2019improving} leveraged transfer learning and five augmentation methods, including rotation, 3D rotation, scaling, jittering, and permutation. 
Their study showcased performance enhancements in two out of three CNN models, emphasizing the efficacy of specific augmentation combinations.
Chung \etal \cite{chung2019sensor} focused on optimizing sensor positions and sensor fusion to classify daily activities, utilizing the jittering method for data augmentation. 

Cheng \etal \cite{cheng2023learning} introduced a contrastive supervision approach for time series data augmentations, emphasizing the importance of learning hierarchical augmentation invariance across different depths of neural networks. 
Their work highlighted that deeply supervised learning, coupled with contrastive losses at intermediate layers, could prevent information loss induced by augmentation.

Zhou \etal \cite{zhou2024autoaughar} recently introduced a two-stage, gradient-based data augmentation framework, AutoAugHAR, to address cross-subject generalization challenges in HAR. Unlike traditional augmentation methods, AutoAugHAR optimizes augmentation operations directly during model training. This model-agnostic framework improves dataset representativeness and robustness without significantly increasing training time and computational costs. 

Despite the merits of data augmentation, it is acknowledged that transforming entire signals may compromise label validity and preservation. 
Kim \etal \cite{kim2021label} raised concerns about changing labels when augmenting sensor signals, leading to potential performance degradation due to the similarity of augmented signals to other activity classes. 
Therefore, data augmentation methods are not enough to improve the performance of HAR and it is necessary to generate sensor data by estimating the data distributions precisely.

\subsection{Generating Virtual Sensor Data from Video}
Efforts to bridge the gap between video and Inertial Measurement Unit (IMU) data have gained prominence, with recent works addressing the challenge of translating video data into IMU representations \cite{rey2019let, rey2020yet, kwon2020imutube, kwon2021complex}. 
These works employ generative methods \cite{rey2019let, fortes2021translating} and trajectory-based approaches \cite{kwon2020imutube, kwon2021approaching, xiao2021deep} to extract (virtual) IMU data from videos, expanding the applicability of sensor-based HAR beyond traditional IMU-equipped scenarios. 
An overview of the approaches is shown in \cref{tab:VidGenSens}.

Virtual IMU data generation has emerged as a viable solution to overcome data scarcity, with cross-modality transfer approaches \cite{rey2019let, rey2020yet, kwon2020imutube, kwon2021complex, fortes2021translating} being instrumental in extracting virtual IMU data from 2D RGB videos of human activities. 
Such methods not only expand training datasets for motion exercise recognition but also enable the construction of personalized HAR systems tailored to individual user needs \cite{xia2022virtual}. 
Utilizing virtual IMU data enhances the accuracy and robustness of HAR models across diverse application domains.

Generative methods \cite{rey2019let, fortes2021translating} leverage machine learning to learn functions capable of deriving IMU data directly from videos. 
In contrast, trajectory-based methods, exemplified by \cite{kwon2020imutube, kwon2021approaching, xiao2021deep}, initially determine 3D joint positions from videos and then utilize forward kinematics to estimate joint orientations. 
The obtained orientations enable the transformation of 3D joint positions into the IMU's frame-of-reference, facilitating the computation of acceleration and angular velocity. 
Notably, the majority of these endeavors have been geared towards human activity recognition tasks, where synthetic IMU data is generated for multiple body joints, allowing for compensation of errors in the estimation of one joint by another.

\begin{table}[!t]
    \centering
    \caption{
    Summary of works which study generation of virtual IMU data from videos for HAR.
    }
    \resizebox{\textwidth}{!}{%
    \begin{tabular}{c P{2cm} c P{4cm} P{5cm}}
        \toprule
         Approach & Category & Input Source & Outputs & Remarks \\
         \midrule
         Rey \etal \cite{rey2019let, rey2020yet} & Generative & 2D RGB Videos & 2D poses and Direct IMU Signal Estimates  & Extracts IMU data directly from videos to increase data flexibility \\
         % \hline
         \hdashline \noalign{\vskip 0.4ex}
         Kwon \etal \cite{kwon2020imutube} & Trajectory-Based  & 2D RGB Videos & 3D Joint Orientations, IMU Data & Uses adaptive selection, tracking, and data extraction to produce virtual IMU data from videos. \\
         % \hline
         \hdashline \noalign{\vskip 0.4ex}
         Kwon \etal \cite{kwon2021complex} & Trajectory-Based  & 2D RGB Videos & 3D Joint Orientations, IMU Data  & Focuses on deriving joint orientations and IMU frames, compensating for joint-specific errors. \\ 
         % \hline
         \hdashline \noalign{\vskip 0.4ex}
         Rey \etal \cite{fortes2021translating} & Generative  & 2D RGB Videos & Direct IMU Signal Estimates & Generates IMU signals from videos, aiming to reduce reliance on physical IMU sensors. \\ 
         % \hline
         \hdashline \noalign{\vskip 0.4ex}
         Xia \etal \cite{xia2022virtual} & Augmentation Technique & 3D Motions & Virtual Acceleration Data & Utilizes a spring-joint model with 3D motions. \\
         \bottomrule        
    \end{tabular}
    }
    \label{tab:VidGenSens}
    \vspace*{-1em}
\end{table}
    
IMUTube \cite{kwon2020imutube} exemplifies a system designed to extract virtual IMU data from 2D RGB videos using CV  methods, such as pose tracking.
IMUTube operates as a processing pipeline, integrating computer vision, graphics, and machine learning models to convert large-scale video datasets into virtual IMU data suitable for training sensor-based HAR systems. 
Its three main components--adaptive video selection, 3D human motion tracking, and virtual IMU data extraction and calibration--collectively contribute to the generation of high-quality virtual IMU data. 
The system's versatility has been demonstrated in improving model performance when integrating real and virtual IMU data \cite{kwon2020imutube, kwon2021complex}. 
To enhance the quality of virtual IMU data, Xia \etal \cite{xia2022virtual} proposed a spring-joint model to augment the virtual acceleration signal.
Despite the efficacy of systems like IMUTube, challenges remain, particularly concerning the quality of input videos. 
The limitations of vision-based systems are evident when videos exhibit camera ego-motion or include irrelevant scenes, requiring meticulous video selection. 

\subsection{Generative Adversarial Networks and Diffusion Models for Data Generation and Augmentation}

\subsubsection{Generating Sensor Data using GANs}
Generative models are a class of ML algorithms designed to model and generate data that resembles a given dataset. 
They learn the underlying probability distribution of the data, enabling them to generate new, realistic instances that are consistent with the training data. 
These models have been applied for a wide range of applications, including image and text generation, data synthesis, and anomaly detection. 
In the context of HAR, generative models have demonstrated their utility in capturing the temporal and spatial characteristics of human activities through a diverse set of data sources, including wearable sensors. 
These models have the potential to generate meaningful and informative representations of human activities, making them invaluable for both analysis and synthesis of activity data.

The application of generative models in HAR presents a unique set of challenges.
Sensor data can be sparse and noisy. 
Unlike image data, which is typically well-structured, sensor readings are subject to various sources of interference, making it challenging to model and generate accurate sequences. 
Activities are inherently temporal, and sensor data streams often exhibit complex temporal dependencies. 
Generative models need to capture these dependencies to create meaningful and realistic activity sequences. 
Sensor data can have high dimensionality, particularly when multiple sensors are involved. 
Dimensionality reduction and feature engineering are essential to ensure generative models can effectively capture the data's underlying structure. 
In addition, activities can vary significantly between individuals. 
Generating data for different people while maintaining meaningful patterns is a challenging task. 
Furthermore, acquiring labeled training data for generative models in the context of human activity recognition is often costly and time-consuming. 
Collecting accurate ground truth annotations for wearable sensor data can be particularly challenging.

Traditional oversampling techniques, such as the Synthetic Minority Over-sampling Technique (SMOTE) \cite{chawla2002smote}, SVM-SMOTE \cite{nguyen2011borderline}, and the Majority Weighted Minority Oversampling Technique (MWMOTE) \cite{barua2012mwmote}, have been employed to mitigate data scarcity. 
However, these methods were not specifically designed for Human Activity Recognition (HAR) and thus often overlook temporal dependencies and statistical properties inherent in wearable sensor data. 
Consequently, their effectiveness in capturing the intricate distribution of real-world wearable sensor data is limited, leading to synthetic data that may not faithfully represent the complexities of the original data.

In contrast, GAN-based methods have shown promise in generating realistic time-series data, combining unsupervised and supervised training approaches. 
Yao \etal \cite{yao2018sensegan} proposed SenseGAN, a semi-supervised deep learning framework for IoT applications. 
SenseGAN leverages abundant unlabeled sensing data, minimizing the need for manual labeling. 
It employs an adversarial game involving a classifier, a generator, and a discriminator to jointly train and improve performance.
However, this work only focused on simple IoT applications, not on HAR tasks. 
Wang \etal \cite{wang2018sensorygans} introduced SensoryGAN, a GAN-based framework for sensor-based HAR. 
Addressing the challenge of limited sensor data in practical scenarios, the authors propose three activity-special GAN models, trained with the guidance of vanilla GANs, to generate synthetic sensor data. 

TimeGAN \cite{yoon2019time} introduced a method for generating realistic time-series data by combining unsupervised and supervised training. 
It utilized a learned embedding space jointly optimized with adversarial and supervised objectives, ensuring that the generated sequences maintain the temporal dynamics present in the training data. 
Empirical evaluations demonstrate superior performance compared to state-of-the-art benchmarks in terms of similarity and predictive ability across various real and synthetic time-series datasets -- yet with no meaningful improvements for actual HAR tasks. 

ActivityGAN \cite{li2020activitygan}, utilized a convolutional GAN architecture for data augmentation in sensor-based HAR. 
The model consists of a generation component employing one-dimensional convolution and transposed convolution layers, and a discrimination component using two-dimensional convolution networks. 
The study demonstrated the effectiveness of ActivityGAN in generating synthetic data for human activity simulation, presenting visualizations and evaluating the usability of synthetic data in combination with real data for training HAR models.

`Balancing Sensor Data Generative Adversarial Networks` (BSDGAN) \cite{hu2023bsdgan} addresses the issue of imbalanced datasets in HAR. 
It utilizes an autoencoder to initialize training and generated synthetic sensor data for rarely performed activities. 
Experimental results in the paper demonstrated that BSDGAN effectively captures real human activity sensor data features, and the balanced dataset enhances recognition accuracy for activity recognition models deployed on WISDM and UNIMIB datasets.

A limitation common to all aforementioned works is their dataset and label specificity. 
They generate new sensor data based on the available data and labels but lack the capability to simulate data for various sensor placements or target activities. 
This limitation is particularly relevant when utilizing online video repositories to obtain sensor data for diverse activities and placements.
Furthermore, GAN-based methods demand a substantial amount of labeled data for training, often a challenge in wearable sensor-based scenarios. 
Mode collapse and a lack of diversity in generated data are additional concerns that may limit their efficacy in improving HAR model performance. 
Addressing these challenges remains an area for further exploration and refinement in the field of synthetic sensor data generation for HAR.

\subsubsection{Generating Sensor Data using Diffusion Models}

In recent advancements, Diffusion Probabilistic Models (DM) have outperformed GANs in image synthesis, demonstrating superior results in terms of both quality and diversity \cite{Ho2020DDPM, nichol2021improved, Dhariwal2021DMbeatGANs, rombach2022high, Ho2022VideoDM, gu2022vector}. 
Drawing inspiration from non-equilibrium statistical physics, diffusion models gradually eliminate structure in a data distribution through an iterative forward diffusion process. 
Subsequently, a reverse diffusion process is learned, reinstating structure and yielding a flexible data generative model \cite{sohl2015deep}.

Ho \etal introduced a diffusion process represented as a Markov chain, transforming the original data distribution into a Gaussian, and a reverse process learning to generate samples by progressively removing noise using a DL model \cite{Ho2020DDPM}. 
The denoising U-Net architecture serves as a potent model in this context \cite{ronneberger2015u, Ho2020DDPM, Ho2022VideoDM, jolicoeur2020adversarial}. 
Enhancements to denoising performance include the incorporation of residual layers and attention mechanisms  \cite{song2021score, rombach2022high}. 
However, the usability of the diffusion model is hindered by the considerable computational steps required for high-quality sample generation. 
Researchers have explored techniques like discretization optimization, non-Markovian processes, and partial sampling to accelerate the speed of diffusion models \cite{dockhorn2022genie, song2021denoising}.
While diffusion models have found widespread applications in tasks such as image in-painting, 3D shape generation, text generation, or audio synthesis, their adoption for HAR has been relatively limited so far \cite{lugmayr2022repaint, gong2023diffuseq, kong2021diffwave, xu2022geodiff}. 
Notably, Shao \etal applied a diffusion model with a redesigned U-Net for synthetic sensor data generation \cite{shao2023study}.

Zuo \etal \cite{zuo2023unsupervised} addressed the challenge of expensive and hard-to-annotate sensor data by leveraging unlabeled sensor data accessible in real-world scenarios. 
The architecture of their diffusion model conditions the  model on statistical information like mean, standard deviation, Z-score, and skewness. 
By capturing the statistical properties of sensor data, the  model generates synthetic sensor data closely resembling the characteristics of the original data. 
Unlike traditional generative models dependent on class labels, this approach operates without labeled sensor data, making it highly applicable when labeled data is scarce. 
The framework involves two steps: 
\textit{(i)} training the unsupervised statistical feature-guided diffusion model on large amounts of unlabeled sensor data;  and 
\textit{(ii)} training an independent human activity classifier using a combination of limited labeled real data and synthetic data generated by the pre-trained diffusion model. 
This two-step process effectively combines the strengths of unsupervised learning and supervised classification, leading to enhanced HAR performance.

\vspace*{-1em}
\subsection{Generating Sensor Data using Simulations}
While significant progress has been made in generating sensor data from video data, relatively fewer works focus on generating IMU sensor data directly using simulation. 
Simulation can reduce the need for extensive real-world data collection, thereby accelerating the development of motion recognition systems.
One of the pioneering tools in this field is IMUSIM \cite{young2011imusim}, which introduced a virtual IMU sensor system that simulates acceleration, angular velocity, and magnetic data from 3D motion sequences captured by motion capture (MoCap) equipment. 

Building on this, Kang \etal \cite{kang2019towards} utilized Unity to embed animations and extract virtual IMU data to train classifiers capable of recognizing real-world activities such as standing, walking, and jogging. 
Jiang \etal \cite{jiang2021model} utilized OpenSim, an open-source software system for biomechanical modeling, to simulate individuals with diverse physiological characteristics performing various movements to augment the IMU dataset. 
Xia \etal \cite{xia2022virtual} proposed a virtual IMU sensor module with a spring-joint model to generate augmented acceleration signals from 2D video, reducing the cost of training datasets for motion exercise recognition systems. 
CROMOSim \cite{hao2022cromosim} is a cross-modality sensor simulator designed to generate high-fidelity virtual IMU data from motion capture systems or monocular RGB cameras. 
Using a 3D skinned multi-person linear model (SMPL), it simulated sensor data from arbitrary on-body positions and traind a CNN model to map imperfect 3D body pose estimations to IMU data. 
Uhlenberg \etal \cite{uhlenberg2023co} generated synthetic accelerations and angular velocities using a simulation framework, enabling a detailed analysis of gait events. 
Similarly, Tang \etal \cite{tang2024synthetic} employed OpenSim and forward kinematic methods to create a substantial volume of synthetic IMU data for fall detection. 
However, most early approaches relied on advanced MoCap equipment to reconstruct 3D human motion and generate corresponding virtual IMU signals, not to aim to create realistic IMU sensor readings, and gyroscope data is often excluded.

\section{The Potential and Promises of Foundational Models}
\label{sec:future}
The emergence of Large Language Models (LLMs) \cite{touvron2023llama, achiam2023gpt, jiang2023mistral, team2024gemma} and the subsequent transition to combined language/vision models (Large Language Vision Models, LLVMs \cite{liu2024visual, liu2024improved, beyer2024paligemma, achiam2023gpt, bai2023qwen, wang2023cogvlm}) has revolutionized many AI-related research areas and applications.  
The scope of the changes such models bring to AI, which goes far beyond classical language and computer vision tasks, has led to the emergence of the term  ``foundational models'' as a more general designation, which we adopt in this paper. 

On an abstract level, there is currently a lot of discussion in the AI community about the exact capabilities of such models. 
It goes from seeing them as mere ``statistical parrots'' to highly speculative discussions about the ability of such systems to reason and even emerge consciousness-like behavior \cite{baum2023fear}. 
For the purpose of this article, we refrain from such discussion, as the aspects of such models that define their potential for sensor-based HAR are largely uncontroversial. 

At a basic level, language models are trained to replicate word distributions in online texts.  
That may sound as not obviously being useful for anything beyond text processing, especially not for sensor-based HAR. 
However, there are two things to consider:
\begin{enumerate}
    \item People write text to describe their experiences in, and perception of the real world.  
    This means a significant correlation exists between text distribution statistics and the real world. 
    If a certain sequence of words has a high probability, that it is not only likely to be grammatically correct, but also will often be a true (or common) statement about the real world. 
    
    \item Given the type and amount of text the models are trained on, one could argue that the systems are trained on input from a significant portion of humanity.   
\end{enumerate}

As a consequence, LLMs can be considered a \emph{noisy, but extremely comprehensive approximation of a world model based on the experience of a significant proportion of humanity}.  
LLVMs connect such world models to visual representations, allowing for both advanced image representation and generation. 

\begin{figure}
    \centering
    \includegraphics[width=0.45\columnwidth]{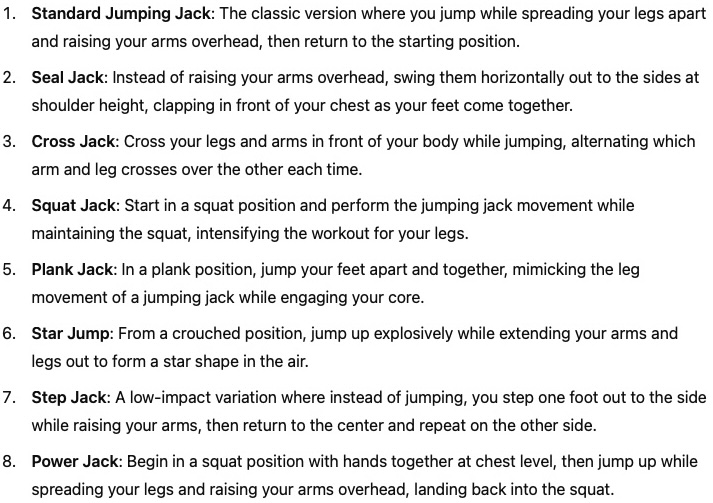}
    \includegraphics[width=0.45\columnwidth]{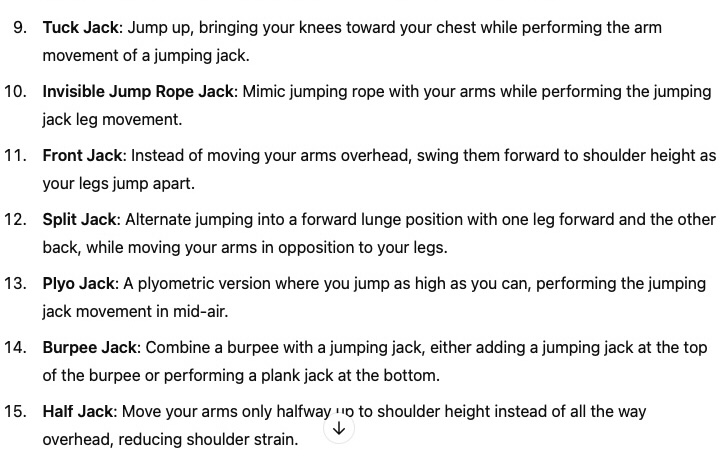}
    \caption{Output of the Open AI O1 model when asked to \texttt{``name a few variations that you could encounter when someone performs jumping jack exercise''}.}
    \label{fig:llmjumpingjack}
    \vspace*{-1em}
\end{figure}

Throughout this paper have we already discussed that the most important problem of sensor-based HAR is the complexity and variability of the real world, which dedicated ``vertical'' training on existing data sets can not reflect. 
Herein lies the significance of LLMs for sensor-based HAR: they are a potential solution to dealing with this variability. 
Thus, for virtually any conceivable activity, the model can provide information about how it can be executed, including in most cases not only the typical way, but often the most of reasonably feasible other ways to execute it (\cref{fig:llmjumpingjack}). 

Furthermore, the majority of models are based on an internal dense vector representation--``embedding'' \cite{liu2024visual, touvron2023llama}--which is analogous to what we have discussed with respect to self-supervised representation learning in \cref{sec:ssl}.  
Such a representation can be used as a ``hook'' to connect sensor data to the world model contained in an LLM.  
That ``hook'' is also what is used by VLLMS to connect images and text, which means that an appropriate embedding can also be used to connect sensor inputs with visual representations. 
In what follows, we will discuss how the three core aspects of the foundational models described above can contribute to sensor-based HAR:  
\begin{enumerate}
    \item The connection to a comprehensive world model that can provide a detailed description of most variations of most activities, including limited characterization of expected sensor signals; 
    
    \item Dense vector space embedding representation of the above world model that can be easily interfaced with HAR-related sensor data representations (in particular ones derived through self-supervised learning);
    
    \item An already existing connection between textual and visual representations of the world model in VLLMs.
\end{enumerate}

 \emph{We argue that by leveraging these aspects we will finally be able to bring sensor-based HAR to the level of performance that we today see in areas such as visual scene interpretation or language understanding.}  

Overviews of existing approaches of using some aspects of sensor-based HAR can be found in \cite{shoumi2024leveraging} or \cite{ferrara2024large}.

\subsection{Foundational Models for Data Generation}
Data generation and augmentation have already been discussed in \cref{sec:generation}.  
In abstract terms, the various approaches can be summarized as consisting of two steps: 
\emph{(i)} creation of a representation of the activities for which we want to generate  data for; and 
\emph{(ii)} the translation of that representation into synthetic sensor data. 
The representations that we considered were videos \cite{kwon2020imutube, kwon2021approaching}, kinematic motion descriptions (e.g., skeleton) \cite{xia2022virtual}, various simulation scripts and trained generative models \cite{zuo2023unsupervised, li2020activitygan}.  
There are three things that foundational models add to the mix:

\begin{enumerate}
    \item \textit{The ability to translate between textual and visual representations of activities}:  On one hand, this facilitates  \emph{labeled} data generation from videos that have no or only vague captions, as the missing caption information can be filled in by the model. 
    On the other hand, we can use textual descriptions of activities to generate captioned videos, which is easier than searching for existing material or having to record new videos. 
    
    \item \textit{Code generation abilities of foundational models}: These can be leveraged to translate textual descriptions of activities into simulation scripts which in turn generate the required sensor data. 
    This is crucial to the generation of sensor data for which visual representations may not contain enough information (e.g., pressure sensors on the ground, physiological sensors, etc.). 
    In this context, code generation capabilities can also be used to go from abstract descriptions of expected signal variations for different users (e.g. different body types) and environmental conditions to corresponding simulation variations. 
    This requires much less effort than manually setting up simulations. 
    As a special case, there are systems that can directly generate physical representations (which can be used as a basis for simulation-based data generation)  from texts. 
    The most prominent examples are systems like T2M-GPT \cite{zhang2023generating}, MotionGPT \cite{jiang2023motiongpt}, MotionDiffuse \cite{zhang2022motiondiffuse}, and ReMoDiffuse \cite{zhang2023remodiffuse} that directly generate motion representations from textual descriptions. 
    
    \item \textit{The ability to break down activities and their variants into individual small steps}: including the description of possible variations in the way activities are executed. 
    Combined with the translation between texts and videos/code described above,  this means that starting with just a set of names of activities, tool sets based on appropriately fine-tuned foundational models can automatically generate comprehensive multi-modal datasets that include variations in execution, user characteristics, and environmental conditions.  
\end{enumerate}

In the long run foundational model's ability to translate between text, video, simulation scripts, and motion representations should lead to training data for sensor-based HAR becoming as abundant as text and image data is currently. 
Clearly, there are still issues to be solved. 
In what follows, we describe contemporary works that leverage these capabilities of foundational models for generating data: 

\subsubsection{IMUGPT}
Leng \etal \cite{leng2023generating,leng2024imugpt} first demonstrated how foundational models can be used to generate realistic and useful wearable sensor data. 
They leveraged ChatGPT to first generate sentence descriptions of activities, e.g., walking or jumping, available in annotated HAR datasets. 
The sentences were input to a pre-trained text-to-motion model called T2MGPT \cite{zhang2023generating}, which generates 3D human motion sequences, which are represented as a sequence of joint positions.
The joints' rotation and translation were computed using inverse kinematics \cite{yamane2003natural}, followed by IMUSim \cite{young2011imusim}, which calculates the joints' acceleration and angular velocity. 
As a result, virtual inertial data are extracted from any of the 22 joints. 
IMUSim also adds realistic noise to the generated data. 

The ``virtual' IMU data has a domain gap to real IMU data collected by placing sensors on participants, due to differences in coordinate systems, sensor orientations, ground forces, etc.\ \cite{leng2023generating}. 
To mitigate this gap, the distribution mapping scheme used in IMUTube \cite{kwon2020imutube} is employed, which uses a few minutes of target data for mapping. 
Training solely on the virtual IMU data performed slightly worse than using real data, but a combination of real and virtual IMU data is clearly more advantageous. 
IMUGPT2.0 \cite{leng2024imugpt} is an extension to this setup, and introduces a motion filter (to filter out irrelevant generated sequences) and metrics to evaluate the diversity of generated data, resulting in a 50\% reduction in the effort necessary for generating virtual IMU data.

\subsubsection{``Text me the data''}
Beyond IMUs, Ray \etal \cite{ray2024text} employ a similar pipeline for generating \textit{pressure map sequences} from text descriptions.
%(\cref{fig:textmethe}. 
GPT-4 \cite{achiam2023gpt} is employed to generate sentences from activity labels, which are used by T2MGPT \cite{zhang2023generating} to produce 3D pose sequences. 
Subsequently, SinMDM \cite{raab2023single} diversifies the generated sequences, followed by conversion into SMPL format \cite{loper2023smpl}.
The pressure sequences are generated using PresSim \cite{ray2023pressim}, which uses volumetric pose as input. 
The utility of generated data for HAR is validated on a newly collected dataset, and a combination of real and synthetic data substantially outperforms using real pressure data only.

\iffalse
\begin{figure}
    \centering
    \includegraphics[width=0.75\columnwidth]{Figures/textmethedata.jpg}
    \caption{\cite{ray2024text}
    \harish{can be removed}}
    \label{fig:textmethe}data
\end{figure}
\fi

\subsection{Reasoning Based on Semantic Information from Foundational Models}
Then, the notion of using background knowledge to compensate for the ambivalence of sensor-based information has been around since the early days of HAR research \cite{bobick1997movement}, initially in vision-based approaches \cite{minnen2003expectation}, with later adoption in sensor-based work (e.g. \cite{patterson2005fine}) that have shown how reasoning about interaction with objects recognized through wearable RFID  can be used for reliable recognition).  The idea is that the possibility/probability of performing a given activity at a given time is closely related to contextual information such as (semantic) location, objects involved as well as previously executed activities.  Thus for example in \cite{bahle2014recognizing} it has been shown that complex activities in a nursing scenario can be recognized using a mobile phone in a pocket of a loose nurse's coat (which is a very poor source of information) by integrating high-level semantic knowledge about the activities in the recognition process. 

Another direction at using semantic information has been the decomposition of more complex activities into basic actions which can be fairly simply recognized from sensors. This way the HAR problem is reduced to the recognition of these simple components and the semantic rules for their composition. The general idea has been reflected in a broad range of work on hierarchical HAR such as, e.g.,  \cite{liao2005location}. 

While such approaches have achieved interesting results in many experiments, they have only had limited impact in the broader field of sensor-based HAR, especially since the emergence of Deep Learning and self-supervised representation learning as described in previous sections. The main reason is the effort involved in manually specifying the relevant semantic relationships. This includes in particular the necessity to cover not just the typical case, but also all sorts of anomalies that may occur in “in the wild” settings.   The effort involved in the manual coding of semantic relationships has motivated the use of ontologies as source of information \cite{chen2009ontology, boovaraghavan2023tao}.  However, the scope of such ontologies is also limited, in particular when it comes to covering the entire breadth of less common situations.

\begin{figure}
    \centering
    \includegraphics[width=0.45\columnwidth]{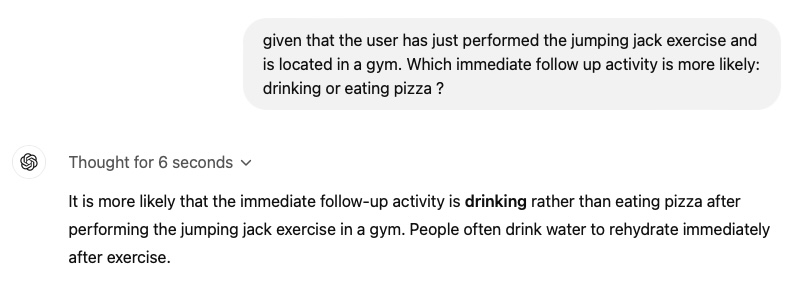}
    \includegraphics[width=0.45\columnwidth]{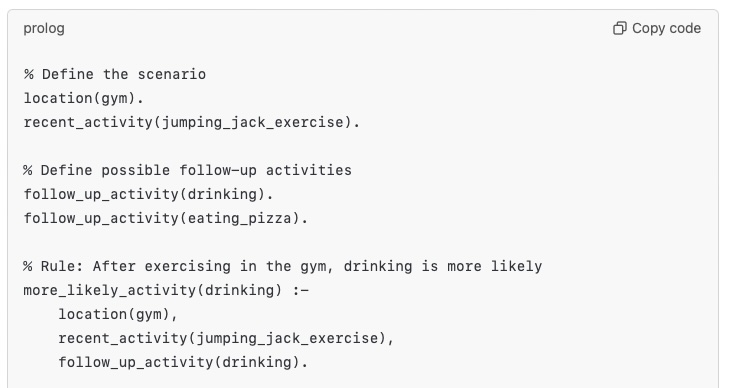}
    \caption{Output of the Open AI O1 model when asked about the probability of different activities as follow up on executing jumping jacks in a gym setting and later asked to convert the answer to Prolog code.}
    \label{fig:followupjack}
    \vspace*{-1em}
\end{figure}

This is where the comprehensive world model contained in foundational models can be a core component of a solution. As already described, for a vast majority of possible situations and activities, the model is likely to contain information on most ways to execute them and the associated situational context. 
The general idea is not new:  in \cite{perkowitz2004mining} it has been proposed to mine activity descriptions from the web.  The difference lies in the scope and depth of the information that can be retrieved as well as the ease with which it can be acquired.  
This is illustrated in Figure \ref{fig:followupjack}.  In addition, a variety of tools exist for converting the model output into some sort of formal representation, particularly code in various computer languages (see \cite{fan2023large} for an overview). Thus, for example in \cite{yang2024arithmetic} the generation of LLM output as Prolog program has been demonstrated as a means of solving arithmetic tasks. A similar approach could be taken to specifying and solving logical constraints on activity recognition (as also illustrated in Figure \ref{fig:followupjack}).  

The use of foundational models, in particular LLMs as source of semantic information to supplement sensor-based HAR is rapidly gaining momentum \cite{hota2024evaluating}. 
In a preliminary study   \cite{arrotta2024contextgpt} the authors have shown that the information provided by LLMs is comparable in terms of usefulness for HAR tasks to information that could be extracted from a dedicated ontology.  
In  \cite{civitarese2024large} this capability of LLMs is used for zero-shot activity recognition in a smart home. The approach is based on the fact that the sensors deployed in the specific location correspond to semantically meaningful events such as user location (OnCouch, NearStove) and device activity (MicrowaveOn, FridgeOn). 
The sequences of such events are then converted into a text prompt, augmented with additional information such as time of day, previous activities, and a list of potential activities to recognize. 
They demonstrate recognition rates that are close to what can be achieved with supervised training on the raw sensor data.   
A similar approach is described in  \cite{chen2024towards}.
Further, TDOST \cite{thukral2024layout} recently demonstrated how textual descriptions of ambient sensor triggers leads to more effective transfer learning across smart home layouts.

\begin{figure}
    \centering
    \includegraphics[width=0.35\columnwidth]{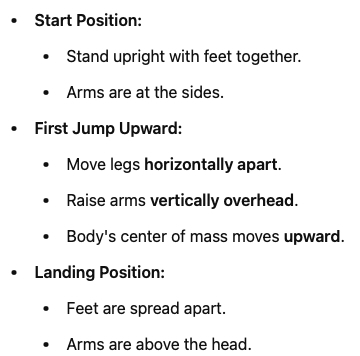}
    \includegraphics[width=0.45\columnwidth]{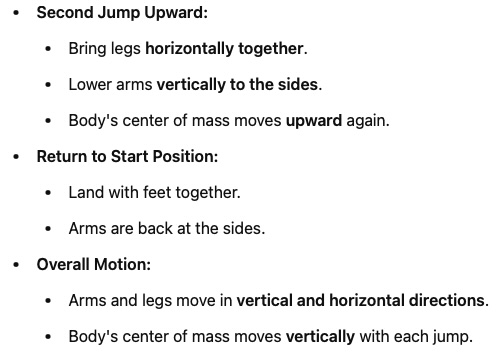}
    \caption{Output of the Open AI O1 model when asked  ``describe the jumping jack exercise in few short sentences by only using vertical and horizontal arms and legs motions and the motion of the body center of mass'' } 
    \label{fig:jumping}
    \vspace*{-1em}
\end{figure}
Using such an approach for sensors such as IMUs is more complex and depends and depends on the ability to translate sensor data windows into semantically meaningful concepts. One possibility would be to identify basic movements of the limbs (up, down sidewards) which LLMs can translate into more complex activities as illustrated in Figure \ref{fig:jumping}

\subsection{Multi-modal Representation Alignment through and with Foundational Models}
Another approach for incorporating knowledge from foundational models is via alignment between data from modalities, e.g., images/videos \cite{radford2021learning, xu2021videoclip, ma2022x}, speech and audio \cite{elizalde2023clap, guzhov2022audioclip, wu2022wav2clip, wu2023large}, or sensor data \cite{haresamudram2024limitations, xia2024ts2act}, with natural language descriptions the data. 
For example, in computer vision, such alignment is performed between images and text captions describing the contents of the image. 
The idea is that the rich expressivity of natural language can describe and supervise a wide variety of concepts -- beyond simple class labels.

For instance, traditional supervised training is performed through class indices ($0,...,N-1$ for $N$ classes), which does not incorporate any additional information about the classes.
Therefore, such classifiers are constrained to predicting any input as one of the $N$ classes.
On the other hand, descriptions of input data (e.g., captions of images) can contain a lot of auxiliary information, thereby allowing a wider range of concepts to be utilized.
For example, a caption for an image of the class `Walking' (having class ID $k \in N$) might be \texttt{``In my new red shirt walking Pepper, my Labrador!''}, which contains additional information about the scene, including the presence of a dog, its name, and its breed, as well as the color of the shirt. 

Alignment between images and captions was popularized by the Contrastive Language-Image Pre-training (CLIP) method \cite{radford2021learning}, which curated a large-scale dataset by crawling the internet for images and their captions, resulting in 400M image-text pairs.
These pairs were first used for cross-modal contrastive pre-training, where the task was to identify which caption was a match for each image. 
Subsequently, the class names in downstream datasets were converted into sentences using simple text templates, e.g., \texttt{``This is a photo of $\{class\_{name}\}$}. 
The class prediction was assigned to the class sentence whose embedding is the closest match to the image embedding. 
This amounts to \textit{zero-shot prediction} of classes, as no additional classifier training is needed (unlike other paradigms such as self-supervised learning).
Furthermore, unseen classes can be predicted, as the predictions are based on similarity to embeddings from class sentences which are extracted from pre-trained language models.
This setup is therefore largely \textit{plug-and-play}, leading to the increasing popularity of the CLIP-style setup and its variants, e.g., UniCL \cite{yang2022unified} and SLIP \cite{mu2022slip}. 
However, the cornerstone for effective performance is the availability of \textit{diverse and rich text descriptions} accompanying \textit{large-scale data} \cite{fan2023improving}. 

\paragraph{Sensor-Language Alignment} 
Moon \etal first applied this paradigm to wearable sensors, through IMU2CLIP \cite{moon2022imu2clip}, which uses the Ego4d dataset \cite{grauman2022ego4d} for training, containing a head-mounted IMU.
TENT \cite{zhou2023tent} aligns text with IoT sensors such as Radar and LIDAR, along with video. 
Recently, Xia \etal introduced Ts2Act \cite{xia2024ts2act}, which performs this alignment between windows of sensor data and \textit{images of activities} obtained from the internet.
As the images are encoded using a pre-trained CLIP image encoder \cite{radford2021learning}, class/activity sentences can be utilized for classification. 
Therefore, this approach connects the sensor-vision-language modalities into a common embedding space.
Similarly, ImageBind \cite{girdhar2023imagebind} connects six modalities into a common embedding space, using pairwise training between vision (images/videos) and other modalities (IMUs, audio, text, and depth and heat data).
The key finding is that co-occurring data from all modalities are not needed for learning a joint embedding space, rather, pairwise aligning of data from modalities with a bridge modality--in this case, vision--leads to \textit{emergent alignment across all modalities.} 
It results in highly interesting capabilities such as embedding space arithmetic, and cross-modal retrieval of samples across modalities, even if they were not pre-trained together originally (e.g., text and audio).
Crucially, however, many of these works perform training and evaluation of sensor-language models on different splits of the same dataset, i.e., between the train-test splits. 
Therefore, the wearable sensors/devices and their locations, and recording conditions all remain largely similar, which contributes to effective performance.

However, the established practice for these models involves cross-modal contrastive pre-training followed by zero-shot prediction on a \textit{collection of diverse target datasets}, which presents a more well-rounded view of their performance. 
Recently, Haresamudram \etal \cite{haresamudram2024limitations} demonstrated that such a setup is highly challenging when training for wearables applications, due to two factors:
\textit{ (i)} sensor heterogeneity -- where diverse sensors result in data with very different distributions, rendering zero shot prediction difficult; and 
\textit{(ii)} lack of rich, detailed descriptions of activities -- most wearable datasets contain class labels only, and in a minority of cases, demographic information, leading to poor alignment between sensor and language modalities.
Consequently, producing plug-and-play sensor-language models remains an unsolved challenge. 
Yet, Haresamudram \etal \cite{haresamudram2024limitations} show that the drop in performance relative to end-to-end training and self-supervision can be reduced by adapting some layers of the network with target data, and through text augmentation. 

Apart from cross-modal alignment, there is also exploratory work for fine-tuning LLMs for HAR purposes.
Imran \etal \cite{imran2024llasa} developed LLaSA, sensor aware question-answering (QA) model combining the LlaMa model \cite{touvron2023llama} with LIMU-BERT \cite{xu2021limu}, and released two datasets, containing IMU data narrations and question-answer pairs, respectively. 
In a similar vein, SensorLLM \cite{li2024sensorllm} sets up a two-stage process in which an LLM is used to align sensor readings with automatically generated text, followed by the recognition of activities. 

Currently, the biggest challenge in developing these models is the lack of a large-scale sensor dataset containing rich descriptions of movements and activities. 
Existing works utilize data from the head-mounted IMU of Ego4D \cite{grauman2022ego4d}, which are unsuited for typical wearable applications at the wrist/waist or use different splits of the same dataset for training and evaluation evaluating performance when there are diverging data distributions.
Therefore, the release of a large-scale datatset, either collected from humans or generated `virtually' can help kickstart research into these techniques, paving the way for multi-modal foundational models integrating wearable sensor data as well.

\subsection{Foundational Models for Time Series Analysis} 

Looking beyond works in re-purposing language and vision foundation models for human activity recognition, significant efforts have been dedicated by researchers to developing foundational models for time-series data. This discussion is relevant to the area of human activity recognition since sensor data can be viewed as time series due to its temporal nature.

TimesFM \cite{das2023decoder} is a time-series forecasting foundation model pretrained on a large corpus of Internet time-series data, including Google Trends and Wiki Pageviews, at the scale of 100 billion data points. The model is trained with the task of point forecasting, predicting directly the values of future time steps. The authors demonstrated that with a relatively small model size (200 million parameters) and a much lower amount of training data when compared to large language models, it remains possible to train a foundation model that achieves a high performance in a wide range of time-series forecasting tasks. This is an inspiring example for the future development of foundation models for human activity recognition: a specialized foundation model could be developed at a fraction of the costs of training large language models.

Chronos \cite{ansari2024chronos} is a family of time-series foundation models trained on a large collection of publicly available datasets. Similar to TimesFM \cite{das2023decoder}, Chronos offers a framework for training an effective zero-shot time-series forecaster without relying on large model sizes (20M to 700M parameters). In this work, the authors proposed to frame the time-series problem as a token prediction task, where time-series values are quantized and converted into tokens, and the model adopts the architecture of large language models for token prediction. This establishes a workaround for the problem of modeling numeric values using token-based transformer networks, in which encoding numeric values as plain-text ASCII strings can hinder the learning process.

MOMENT \cite{goswami2024moment}, is another family of open-source time-series foundation models. The authors argued that training large time series models is often challenging due to the lack of an established collection of datasets and evaluation benchmarks, and the difficulty in handling different characteristics of datasets. In this work, these are addressed by training the models on a collection of publicly available datasets, called Time Series Pile, with different techniques such as sub-sampling and padding to handle time series of different lengths. Masked reconstruction is used as the learning objective, and the authors observed that training from scratch with randomized weights allows the models to converge to a lower training loss when compared to warm starting the model training with LLM weights. With a wide selection of tasks and metrics, MOMENT observes improved performance with parameter scaling (from 40M to 385M parameters).

Although the main focus of these works is often general-purpose time-series analysis, and the applicability of their findings in human activity recognition is yet to be verified, these works provide a good vision of how foundational models for human activity recognition can be: that training with a well-selected set of datasets and modest model size could offer much more practical and effective solutions for this mobile sensing task.

\section{Discussion and Conclusion}
\label{sec:conclusion}
The popularity of specific solutions/techniques for sensor-based HAR ebbs and flows. 
From hand-crafted metrics to \textit{learning} useful representations from the data itself, the prevailing paradigm for recognizing activities has changed over time.
The Bulling tutorial \cite{bulling2014tutorial} formulated the ARC with manually engineered features, breaking down the HAR process into distinct steps.
Subsequently, the increase in dataset sizes coupled with the proliferation of cheap and large quantities of computing fueled the adoption of Deep Learning for HAR as well. 
This was the first paradigm shift -- a transition from manual feature engineering to automatically learning relevant features from \textit{annotated data}. 
This resulted in substantial improvements over erstwhile feature featuring, becoming the de facto solution for HAR. 
Combinations of convolutional, recurrent, and attention-based models were explored for this task.
However, the prevailing practice in other domains, e.g., computer vision, of applying hundreds of layers was not possible due to considerably smaller sizes of sensor datasets. In addition, deep CNNs reflect the way information is represented in images: a hierarchy of spatially local structures. Unfortunately such hierarchy is not obvious in sensor data‚ even if it is converted to ‘fake images'. As a consequence the impact of the deep CNN revolution was much less pronounced in sensor based HAR than in computer vision.

Large-scale data collection efforts by the wearables community, e.g., the GLOBEM dataset \cite{xu2023globem} and the UK Biobank project \cite{willetts2018statistical, doherty2017large}, demonstrated the relative ease of collecting wearable movement in the wild, albeit without annotations. 
Such data, coupled with the increasing popularity of self-supervised representation learning in the machine learning community, led to the exploration of \textit{self-supervised methods} by 2019, with the introduction of Multi-task self-supervision \cite{saeed2019multi}. 
A number of papers followed, which introduced novel pretext tasks or adapted and adopted existing ones from other modalities to suit sensor data. 
These methods excelled in situations of label scarcity (which are all too common in wearable computing), as the learned encoder weights were frozen and only the classifier was updated with a few seconds of labeled data / activity. 
Their performance was often comparable to, if not exceeding end-to-end training. 
This was the second paradigm shift, with the community adopting this \textit{`pretrain-then-finetune'} setup.
Currently, self-supervised representation learning is the predominant approach for sensor-based HAR. 

While originally adopted to facilitate self supervised pre-training, the notion of training representations rather than directly training downstream task is also increasingly being exploited to facilitate novel ways of including additional knowledge in the training process. 
The idea is to encode such knowledge in the structure of the embedding space which is typically achieved by some version of contrastive or adversarial learning. 
This allows for example knowledge from sensors available only during training \cite{fortes2022learning}, or an abstract understanding of which aspects of the data are more or less relevant (e.g. user independence \cite{suh2023tasked}) to be encoded in the representation. 

The release of increasingly capable foundational models has a number of implications for sensor-based HAR, as the embedded world knowledge can be used, e.g., to generate `virtual' data and to perform multi-modal alignment with sensor streams. 
There are many promising efforts in this direction, and the field seems to be poised for its third paradigm shift that may finally solve the problem. 
At this pivotal movement, we reflect on sensor-based HAR as a whole, and trace its early days with feature engineering, to learning supervised and self-supervised representations from data, and finally, to what comes next, which involves leveraging foundational models in some capacity for sensor-based HAR, and other related applications.

\begin{acks}
 
Sungho Suh and Paul Lukowicz were partially supported by the European Union’s Horizon 2020 Program under grant no. 952026 HumanE-AI-Net project.
Harish Haresamudram and Thomas Ploetz were partially supported by the NSF grant IS-2112633 and a grant from Optum. 
 
\end{acks}

\bibliographystyle{ACM-Reference-Format}
\bibliography{refs}

\begin{appendix}
\section{Walk-Through Examples of Sensor-Based Human Activity Recognition}

We split the tutorial into three modules, covering different types of representations used for recognizing activities: 
\begin{enumerate}
    \item Distribution-based features, i.e., ECDF \cite{hammerla2013preserving} as an example of features used in traditional HAR;
    \item Conv. classifier \cite{haresamudram2022assessing}, to examine the performance of supervised Deep Learning; and
    \item SimCLR \cite{tang2020exploring, chen2020simple}, for measuring the effectiveness of a highly effective self-supervised method.
\end{enumerate}
In what follows, we present selected snippets of the code that accompanies our tutorial for the recognition of activities with these features, placing emphasis on the major steps rather than low-level details.
The primary metric of evaluation is the mean (or unweighted) F1-score, as it is more resistant to class imbalance, which commonly occurs in wearable datasets \cite{plotz2021applying}.

\paragraph{Dataset:} We utilize the \textit{Motionsense dataset} \cite{malekzadeh2018protecting}\footnote{\href{https://github.com/mmalekzadeh/motion-sense}{https://github.com/mmalekzadeh/motion-sense}}.
We utilize \textit{accelerometer data only}, to reduce the training times for the tutorial.
Yet, the code can be easily extended to work with additional sensors such as gyroscopes.

\begin{figure}[h]
    \centering
    \includegraphics[width=0.6\linewidth]{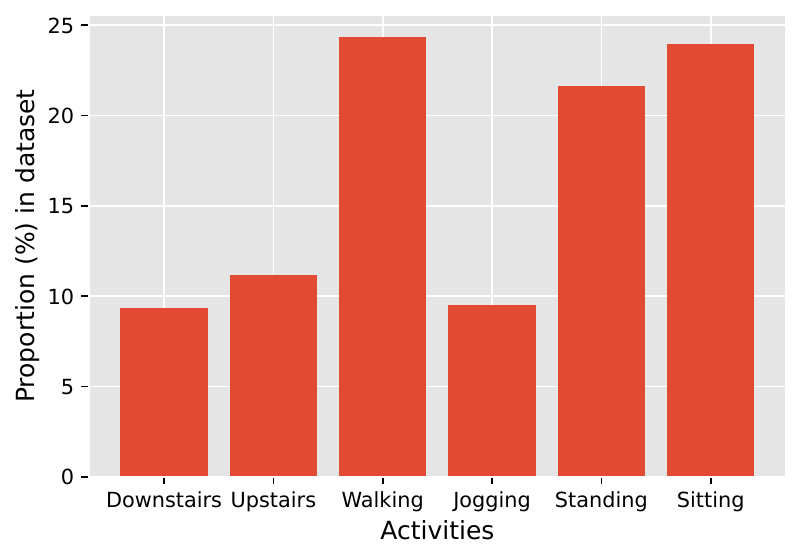}
    \caption{Class composition of the Motionsense dataset.}
    \label{fig:motionsense_classes}
\end{figure}

Motionsense contains six locomotion-style activities from 24 participants, namely, walking downstairs, walking upstairs, walking, jogging, standing, and sitting. 
In Fig.\ \ref{fig:motionsense_classes} we visualize the class composition of the dataset, and note that some classes such as walking, standing, and sitting appear more frequently than others, thereby indicating the presence of class imbalance.

\paragraph{Code:} The code is available in the following Git repository (anonymized for review purposes): \href{https://github.com/submissionimwut/IMWUT\_submission/tree/main}{link}.\footnote{\href{https://github.com/submissionimwut/IMWUT\_submission/tree/main}{https://github.com/submissionimwut/IMWUT\_submission/tree/main}}
It needs to be cloned to work with the tutorial locally. 
Alternatively, the accompanying Jupyter notebook goes through the tutorial step-by-step, along with other helpful visualizations: \href{https://github.com/submissionimwut/IMWUT\_submission/blob/main/Tutorial\_Notebook.ipynb}{link}.\footnote{\href{https://github.com/submissionimwut/IMWUT\_submission/blob/main/Tutorial\_Notebook.ipynb}{https://github.com/submissionimwut/IMWUT\_submission/blob/main/Tutorial\_Notebook.ipynb}}
The slides are available here: \href{https://drive.google.com/file/d/1Z7RgsVKuiuUSfv7DjrTrAelBLSexqynB/view?usp=sharing}{link}.\footnote{\href{https://drive.google.com/file/d/1Z7RgsVKuiuUSfv7DjrTrAelBLSexqynB/view?usp=sharing}{https://drive.google.com/file/d/1Z7RgsVKuiuUSfv7DjrTrAelBLSexqynB/view?usp=sharing}}

\subsection{Baseline: Traditional Human Activity Recognition with ECDF features}
\label{sec:traditional_har}
As detailed in Sec.\ \ref{sec:history}, the Activity Recognition Chain (ARC) \cite{bulling2014tutorial} was employed for traditional HAR, comprising of five steps: \textit{(i)} data collection; \textit{(ii)} pre-processing; \textit{(iii)} segmentation; \textit{(iv)} feature extraction; and \textit{(v)} classification.
In the code snippets below, we focus on steps (ii)-(v), as data have already been collected for public datasets.

\subsubsection{Data Pre-processing}
It typically involves data cleaning, normalization, filtering, etc. -- steps that are often required to clean and prepare the data for classification.
In our example, we first read in the relevant files from the dataset.
Many public datasets are released in the CSV format and each file can contain a different activity, a different participant, or both.
Then, we partition the dataset into training / validation / test splits by first randomly sampling 20\% of the participants to be the test split.
Of the remaining participants, we once again sample 20\% to form the validation set, where as the rest comprise the train split. 

As Motionsense contains 24 participants, this process results in $(15, 4, 5)$ users for train / validation / test, respectively. 
Subsequently, we perform z-score normalization on the train set, i.e., the train data are normalized to have zero mean and unit variance.
The normalization parameters are also used to normalize the validation and test splits. 
By running the code below, we obtain a dictionary called \colorbox{lightgraybackground}{\texttt{processed}}, containing the train / validation / test streams of sensor data, along with associated annotations.
This dictionary is then used for sliding window segmentation.

\begin{minted}
[
numbersep=3pt,
frame=lines,
framesep=2mm,
baselinestretch=1.2,
bgcolor=lightgraybackground,
linenos
]
{python}
# Obtaining the processed data
processed = prepare.prepare_data(args)
\end{minted}

\subsubsection{Segmentation}
We apply the sliding window process to segment the contiguous streams of sensor data into windows (i.e., \colorbox{lightgraybackground}{\texttt{segmented\_data}}).
Here, we utilize a window size of two seconds, with an overlap of one second.
\begin{minted}
[
numbersep=3pt,
frame=lines,
framesep=2mm,
baselinestretch=1.2,
bgcolor=lightgraybackground,
linenos
]
{python}
# Obtaining the segmented data
segmented_data = ecdf.generate_windowed_data(processed=processed)
\end{minted}

\subsubsection{Feature Extraction}
Distribution-based features are extracted for each window in the dataset (across splits).
For ECDF, the number of quantiles, i.e., the number of components is a hyperparameter, with the optimal number depending on the activities under study \cite{kwon2018adding}.
Here, we utilize 25 quantiles for extracting features, resulting in a feature size of 77 per window. 
\begin{minted}
[
numbersep=3pt,
frame=lines,
framesep=2mm,
baselinestretch=1.2,
bgcolor=lightgraybackground,
linenos
]
{python}
# Computing the ECDF features
ecdf_features = ecdf.compute_ecdf_features(segmented_data=segmented_data)
\end{minted}

Finally, we train a Random Forest (RF) classifier using the extracted features.
The code below also prints out the performance on each of the splits. 
In our runs, we obtain a test set F1-score of \textbf{81.84\%}, whereas the performance on the train set is 100\%, indicating potential overfitting.

\begin{minted}
[
numbersep=3pt,
frame=lines,
framesep=2mm,
baselinestretch=1.2,
bgcolor=lightgraybackground,
breaklines,
linenos
]
{python}
# Training the RF classifier
trained_classifier, log_ecdf = ecdf.train_rf_classifier(ecdf=ecdf_features, segmented_data=segmented_data)
\end{minted}

\subsection{HAR with Supervised Deep Learning: Convolutional Classifier}
\label{sec:sup_conv_classifier}
We use the PyTorch framework \cite{paszke2019pytorch} for implementing the classifier.
Below, we describe how steps in the ARC are accomplished using the framework.

\subsubsection{Segmentation and Data Loading}
In Pytorch, the \colorbox{lightgraybackground}{\texttt{torch.utils.data.DataLoader}} wraps the \colorbox{lightgraybackground}{\texttt{HARDataset}} class, which loads the normalized sensor streams, performs segmentation, and passes individual sensor windows and corresponding labels (based on \colorbox{lightgraybackground}{\texttt{index}}) to the data loader.
We show a snippet below, containing only important steps, with full code on GitHub.

\begin{itemize}
    \item \colorbox{lightgraybackground}{\texttt{load\_dataset()}}: loads the data processed in Sec.\ \ref{sec:traditional_har}, as it has already been normalized and split into train/validation/test sets.
    \item \colorbox{lightgraybackground}{\texttt{opp\_sliding\_window()}}: performs  segmentation of sensor streams into windows.
    Here, we pass the sensor data, the associated labels, the window size, and the overlap to be used for segmentation.
    \item \colorbox{lightgraybackground}{\texttt{load\_har\_dataset()}}: creates data loaders for each split of data, i.e., train/validation, and test. 
    During classifier training, it also outputs batches of data and labels and shuffles samples if required. 
\end{itemize}

\begin{minted}
[
numbersep=3pt,
frame=lines,
framesep=2mm,
baselinestretch=1.2,
bgcolor=lightgraybackground,
breaklines,
linenos
]
{python}
class HARDataset(Dataset):
    def __init__(self, args, phase):
        self.filename = os.path.join(args['root_dir'], args['data_file'])
        # [truncated] 
        
        # Loading the data
        self.data_raw = self.load_dataset(self.filename)

        # Obtaining the segmented data
        self.data, self.labels = \
            opp_sliding_window(
            self.data_raw[phase]['data'], self.data_raw[phase]['labels'],
            args['window'], args['overlap'])
        # [truncated] 

    def load_dataset(self, filename):
        data_raw = joblib.load(filename)
        # [truncated] 

        return data_raw
        

    def __getitem__(self, index):
        data = self.data[index, :, :]
        data = torch.from_numpy(data)

        label = torch.from_numpy(np.asarray(self.labels[index]))
        return data, label

def load_har_dataset(args):
    datasets = {x: HARDataset(args=args, phase=x) for x in
                ['train', 'val', 'test']}
    data_loaders = {x: DataLoader(datasets[x],
                                  batch_size=args['batch_size'],
                                  shuffle=True if x == 'train' else False,
                                  num_workers=0, pin_memory=True) for x in
                    ['train', 'val', 'test']}

    dataset_sizes = {x: len(datasets[x]) for x in ['train', 'val', 'test']}

    return data_loaders, dataset_sizes
\end{minted}

\subsubsection{Classifier Training}
We evaluate the performance of a simple convolutional classifier for recognizing activities in Motionsense. 
It contains two parts: a convolutional encoder and an MLP for classification. 
The encoder's architecture is identical to previous works \cite{haresamudram2022assessing, saeed2019multi, tang2020exploring} and contains three blocks.
Inside each block is a 1D convolutional layer, followed by ReLU activation, and dropout with p=0.2. 
Across blocks, the number of filters is set to (32, 64, 96) and the kernel sizes are (24, 16, 8). 
After the encoder, we employ global max pooling to obtain an embedding which is used for classification using the MLP.
It contains two linear layers of size (1024, $num\_classes$) units, with ReLU activation in between. 
For Motionsense, $num\_classes$ is set to 6, resulting in the architecture shown below:
\begin{minted}
[
numbersep=3pt,
frame=lines,
framesep=2mm,
baselinestretch=1.2,
bgcolor=lightgraybackground,
breaklines,
linenos
]
{python}
Classifier(
  (backbone): Encoder(
    (conv1): ConvBlock(
      (conv): Conv1d(3, 32, kernel_size=(24,), stride=(1,))
      (relu): ReLU()
      (dropout): Dropout(p=0.1, inplace=False)
    )
    (conv2): ConvBlock(
      (conv): Conv1d(32, 64, kernel_size=(16,), stride=(1,))
      (relu): ReLU()
      (dropout): Dropout(p=0.1, inplace=False)
    )
    (conv3): ConvBlock(
      (conv): Conv1d(64, 96, kernel_size=(8,), stride=(1,))
      (relu): ReLU()
      (dropout): Dropout(p=0.1, inplace=False)
    )
  )
  (softmax): Sequential(
    (0): Linear(in_features=96, out_features=1024, bias=True)
    (1): ReLU(inplace=True)
    (2): Linear(in_features=1024, out_features=6, bias=True)
  )
)
\end{minted}

We perform training for 50 epochs with the Adam optimizer.
The learning rate and weight decay are set to $10^{-4}$, with a batch size of $256$, with the learning rate reducing by a factor of 0.8 every 10 epochs.

We utilize the processed data from Sec.\ \ref{sec:traditional_har} for classifier training. 
The data loading and classification loops are wrapped using an overarching function \colorbox{lightgraybackground}{\texttt{evaluate\_with\_classifier()}}, shown below (please refer to the Github repository for detailed code).

\begin{minted}
[
numbersep=3pt,
frame=lines,
framesep=2mm,
baselinestretch=1.2,
bgcolor=lightgraybackground,
breaklines,
linenos
]
{python}
def evaluate_with_classifier(args=None):
    # Load the target data
    data_loaders, dataset_sizes = load_har_dataset(args)
    # [truncated] 

    # Creating the model
    model = Classifier(args).to(args['device'])

    # Optimizer settings
    optimizer = optim.AdamW(model.parameters(), lr=args['learning_rate'],
                            weight_decay=args['weight_decay'])
    scheduler = StepLR(optimizer, step_size=10, gamma=0.8)
    criterion = nn.CrossEntropyLoss()

    for epoch in tqdm(range(0, args['num_epochs'])):
        # Training
        model, optimizer, scheduler = train(model, .....)

        # Validation
        evaluate(model, .....)

        # Evaluating on the test data
        evaluate(model, .....)

        # [truncated] 

    # [truncated] 

    return
\end{minted}

Digging deeper into \colorbox{lightgraybackground}{\texttt{evaluate\_with\_classifier()}}, we see that it contains the following (important) components:
\begin{itemize}
    \item \colorbox{lightgraybackground}{\texttt{load\_har\_dataset()}}: creates the data loaders for each split of the dataset. 
    \item \colorbox{lightgraybackground}{\texttt{model = Classifier(args)}}: creates the Conv. classifier, with the model architecture shown above. 
    \item \colorbox{lightgraybackground}{\texttt{optimizer, scheduler, criterion}}: we utilize the Adam optimizer, with a step learning rate schedule, and train with the Cross Entropy loss.
    \item \colorbox{lightgraybackground}{\texttt{for epoch in tqdm(range(0, args['num\_epochs'])):}}: this is the main training loop, containing the training, validation, and testing loops for each epoch. 
    The performance is logged for further analysis.
\end{itemize}

At the end of the training run, the test set performance is \textbf{85.1\%}, showcasing an improvement of approx.\ 3\% over the ECDF-RF classifier combination.

\subsection{HAR with Self-Supervised Learning: SimCLR}
\begin{figure}[!h]
    \centering
    \includegraphics[width=0.9\columnwidth]{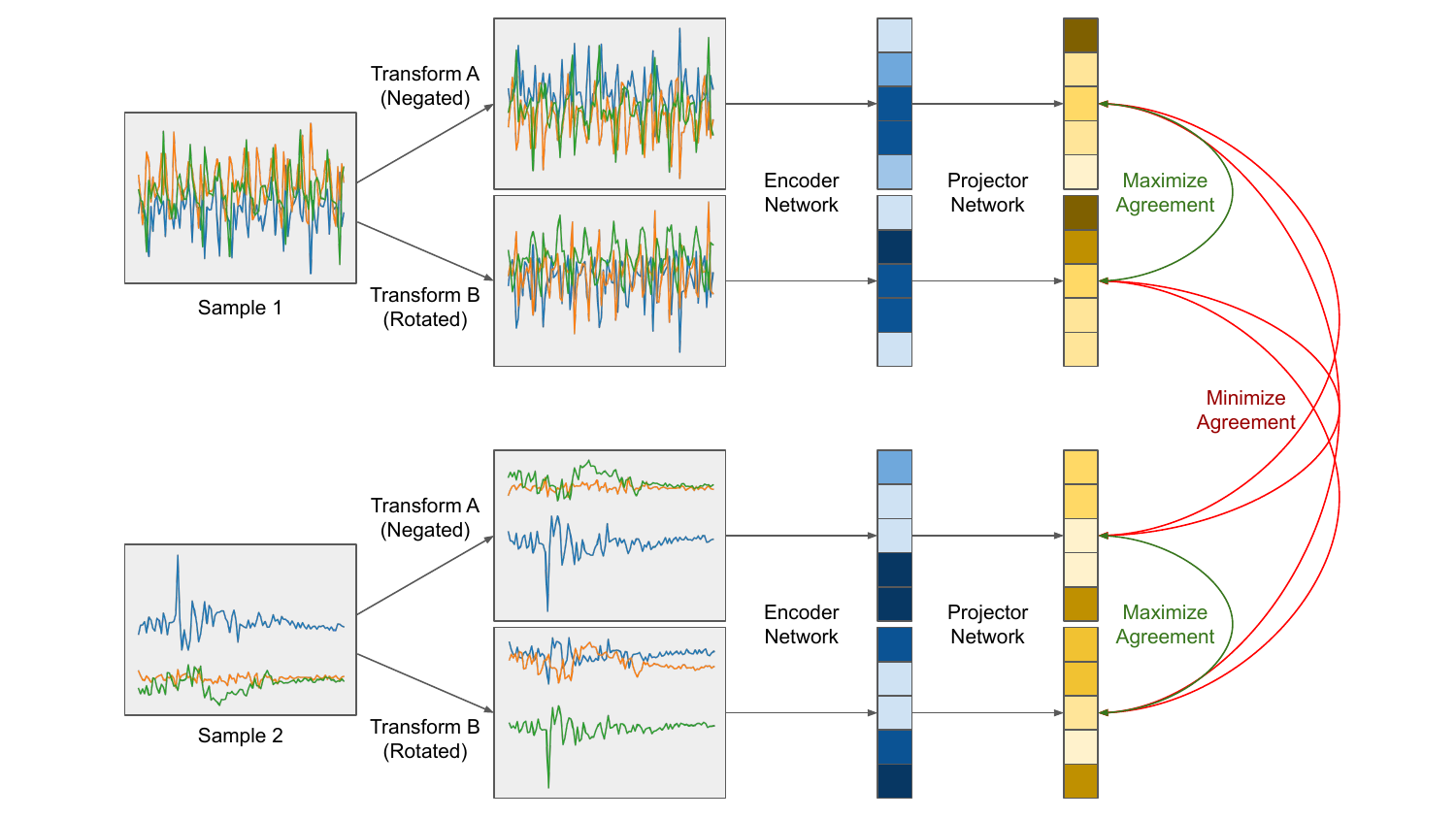}
    \vspace*{-1em}
    \caption{An overview of the SimCLR training pipeline for human activity recognition \cite{tang2020exploring}.}
    \label{fig:simclr_example}
    \vspace*{-1em}
\end{figure}

Self-supervised learning is a two-stage process: (i) pre-training with unlabeled data by solving a pretext task, and (ii) classifying the target activities using representations extracted from the learned encoder weights.
As detailed in Sec.\ \ref{sec:contrastive_adaptations}, the pretext task involves distinguishing between positive and negative samples generated by randomly transforming/augmenting windows of sensor data.
Figure \ref{fig:simclr_example} gives an example of this learning process, in which the agreements of embeddings among different augmented views of the input data are maximized or minimized depending on whether they are positive pairs or negative pairs.

\begin{figure}
    \centering
    \includegraphics[width=0.9\columnwidth]{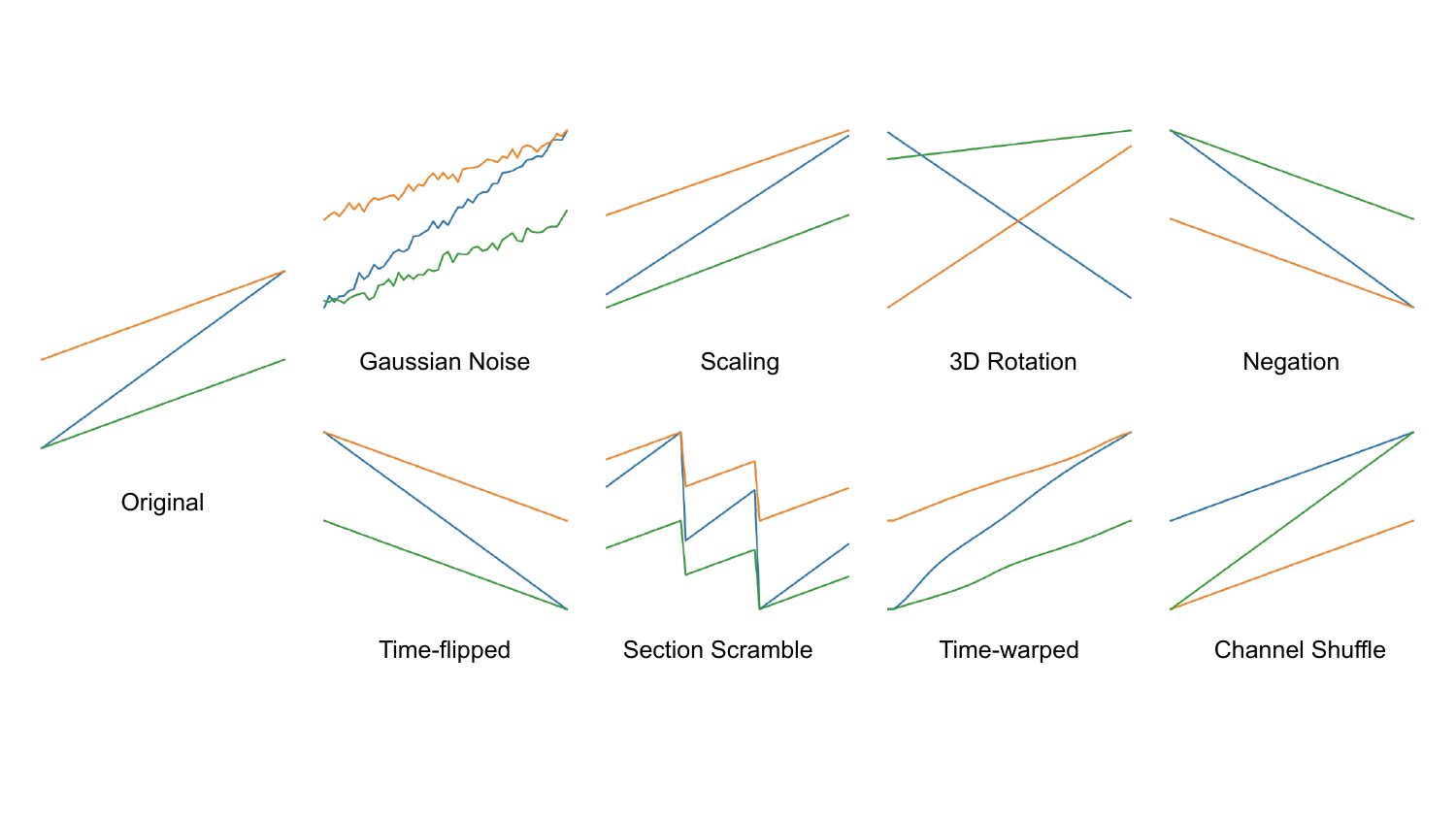}
    \vspace*{-1em}
    \caption{Illustrations of transformation functions used in the SimCLR training pipeline \cite{um2017data, tang2020exploring}.
    }
    \label{fig:simclr_transform}
    \vspace*{-1em}
\end{figure}

The choice of augmentations is key to performance, and in this tutorial, we utilize all pairwise combinations of augmentations introduced by Um \etal \cite{um2017data}. 
We leverage the efficient, vectorized augmentations implemented by Tang \etal \cite{tang2020exploring}, taken from their  \href{https://github.com/iantangc/ContrastiveLearningHAR}{repository}\footnote{https://github.com/iantangc/ContrastiveLearningHAR} (see Figure \ref{fig:simclr_transform}).

The encoder (also called the backbone) contains three 1D convolutional blocks, identical to Tang \etal \cite{tang2020exploring} and Saeed \etal \cite{saeed2019multi} (and in the Conv. classifier discussed previously). 
Each block has a 1D convolutional layer, followed by ReLU and dropout with $p=0.2$.
Further, the number of filters is (32, 64, 96), with kernel sizes of (24, 16, 8) respectively, followed by a global max pooling layer which outputs the embeddings from the encoder.
The projection head is an MLP comprising three linear layers of (256, 128, 50) units and ReLU activation in between.
The architecture is as follows:

\begin{minted}
[
numbersep=3pt,
frame=lines,
framesep=2mm,
baselinestretch=1.2,
bgcolor=lightgraybackground,
breaklines,
linenos
]
{python}
SimCLR(
  (backbone): Encoder(
    (conv1): ConvBlock(
      (conv): Conv1d(3, 32, kernel_size=(24,), stride=(1,))
      (relu): ReLU()
      (dropout): Dropout(p=0.1, inplace=False)
    )
    (conv2): ConvBlock(
      (conv): Conv1d(32, 64, kernel_size=(16,), stride=(1,))
      (relu): ReLU()
      (dropout): Dropout(p=0.1, inplace=False)
    )
    (conv3): ConvBlock(
      (conv): Conv1d(64, 96, kernel_size=(8,), stride=(1,))
      (relu): ReLU()
      (dropout): Dropout(p=0.1, inplace=False)
    )
  )
  (projection_head): Sequential(
    (0): Linear(in_features=96, out_features=256, bias=True)
    (1): ReLU()
    (2): Linear(in_features=256, out_features=128, bias=True)
    (3): ReLU()
    (4): Linear(in_features=128, out_features=50, bias=True)
  )
)
\end{minted}

The pre-training is performed for 50 epochs using the SGD optimizer and the NT-Xent loss with a temperature of $0.1$.
The learning rate is set to $10^{-3}$ with weight decay of $10^{-5}$ and batch size of 1024.
The learning rate is set to follow a cosine annealing schedule, starting with $10^{-3}$ and reaching 0 after 50 epochs.
The pre-training is performed with a wrapper function called \colorbox{lightgraybackground}{\texttt{learn\_model()}}, as shown below:

\begin{minted}
[
numbersep=3pt,
frame=lines,
framesep=2mm,
baselinestretch=1.2,
bgcolor=lightgraybackground,
breaklines,
linenos
]
{python}
def learn_model(args=None):
    # [truncated] 

    # Data loaders
    data_loaders, dataset_sizes = load_har_dataset(args, pretrain=True)

    # Creating the model
    model = SimCLR(args).to(args['device'])

    optimizer = torch.optim.SGD(model.parameters(), lr=args['learning_rate'],
                                weight_decay=args['weight_decay'], momentum=0.9)
    scheduler = torch.optim.lr_scheduler.CosineAnnealingLR(
        optimizer, T_max=args['num_epochs']
    )
    criterion = NTXentLoss(temperature=0.1)

    # List of transformations
    transform_funcs_vectorized = [
        transformations.noise_transform_vectorized,
        transformations.scaling_transform_vectorized,
        transformations.rotation_transform_vectorized,
        transformations.negate_transform_vectorized,
        transformations.time_flip_transform_vectorized,
        transformations.time_warp_transform_low_cost,
        transformations.channel_shuffle_transform_vectorized
    ]

    for epoch in tqdm(range(0, args['num_epochs'])):
        # Training
        model, optimizer = train(model, ....)

        scheduler.step()

        # Evaluating on the validation data
        evaluate(model, ....)

        # [truncated] 

    # [truncated] 

    return
\end{minted}

This function contains the following essential components:
\begin{itemize}
    \item \colorbox{lightgraybackground}{\texttt{load\_har\_dataset()}}: creates the data loaders for each split of the dataset. 
    \item \colorbox{lightgraybackground}{\texttt{model = SimCLR(args)}}: creates the SimCLR model and initializes it with random weights. 
    \item \colorbox{lightgraybackground}{\texttt{optimizer, scheduler, criterion}}: we utilize the SGD optimizer, with a cosine annealing learning rate schedule, and train with the NTXent loss
    \item \colorbox{lightgraybackground}{\texttt{transform\_funcs\_vectorized}}: defines the list of augmentations/transformations to be applied on the sensor data windows, including adding random Gaussian noise, scaling, rotations, sensor channel shuffling, etc.
    \item \colorbox{lightgraybackground}{\texttt{for epoch in tqdm(range(0, args['num\_epochs'])):}}: this is the main pre-training loop, where the contrastive task is solved on both training and validation. 
    This performance is logged for further analysis.
\end{itemize}

\paragraph{SimCLR pre-training}
The \colorbox{lightgraybackground}{\texttt{train()}} function, which implements the contrastive learning objective, is given below:
\begin{minted}
[
numbersep=3pt,
frame=lines,
framesep=2mm,
baselinestretch=1.2,
bgcolor=lightgraybackground,
breaklines,
linenos
]
{python}
def train(model, ... ):
    # [truncated]

    # Iterating over the data
    for inputs, _ in data_loader:
        if len(trans_comb) == 0:
            trans_comb = [i for i in itertools.permutations
            (range(len(transform_funcs_vectorized)), 2)]

        # Getting each transform pair
        i1, i2 = trans_comb.pop()
        t1 = transform_funcs_vectorized[i1]
        t2 = transform_funcs_vectorized[i2]

        # [truncated]

        # Transforming the input batch two-ways
        data_1 = torch.from_numpy(t1(inputs).copy()).float().to(args['device'])
        data_2 = torch.from_numpy(t2(inputs).copy()).float().to(args['device'])

        with torch.set_grad_enabled(True):
            outputs_1 = model(data_1)
            outputs_2 = model(data_2)

            loss = criterion(outputs_1, outputs_2)

            # [truncated]

        # Appending predictions and loss
        # [truncated]
    
    # [truncated]

    return model, optimizer
\end{minted}

Here are the descriptions of the key parts of this function:
\begin{itemize}
    \item \colorbox{lightgraybackground}{\texttt{for inputs, \_ in data\_loader:}}: First we load data from the data loader in batches (\colorbox{lightgraybackground}{\texttt{inputs}}).
    \item \colorbox{lightgraybackground}{\texttt{itertools.permutations(range(len(transform\_funcs\_vectorized)), 2)}}: Here we refresh the pool of transformations functions that are used to augment data. 
    We generate all combinations of pairs of transformations, as defined above.
    \item \colorbox{lightgraybackground}{\texttt{data\_1 = t1(inputs)}}: For each pair of transformation functions \colorbox{lightgraybackground}{\texttt{t1, t2}}, two views of each sample are generated by applying each of these functions separately.
    \item \colorbox{lightgraybackground}{\texttt{outputs\_1 = model(data\_1)}}: We then pass these transformed views of the data through the encoder.
    \item \colorbox{lightgraybackground}{\texttt{loss = criterion(outputs\_1, outputs\_2)}}: By using \colorbox{lightgraybackground}{\texttt{NTXentLoss}} as the \colorbox{lightgraybackground}{\texttt{criterion}}, we calculate the loss by passing the two sets embeddings generated by the encoder.
    \item The NTXent loss function, as adopted in \cite{chen2020simple, tang2020exploring}, calculates the loss by taking the elements with the same index from both views as positives, while all other samples as negatives. These are minimized in an analogous way to the cross-entropy loss for multi-label classification: $\-\log \frac{\exp(\text{sim}(o_i^1, o_i^2)/\tau)}{\sum_{j=1}^{N} 1_{j \neq i}\exp(\text{sim}(o_i^1, o_j^1)/\tau) + \exp(\text{sim}(o_i^1, o_j^2)/\tau)}$, where $o_i^1$ stands for \colorbox{lightgraybackground}{\texttt{outputs\_1[i]}}, $\text{sim}$ is the cosine similarity function, and $\tau$ is the temperature parameter.
\end{itemize}

Once the pre-training is complete, we freeze the learned encoder weights and utilize them for HAR. 
We discard the projection head used during pre-training and instead replace it with an MLP classifier (alternatively, a linear classifier can also be used), and train \textit{only the MLP} to recognize activities. 
The MLP contains two linear layers of (1024, 6) units respectively, with ReLU activation. 
This architecture matches the supervised Conv. classifier from Sec.\ \ref{sec:sup_conv_classifier}, but the encoder remains frozen during HAR. 
The architecture is as follows:
\begin{minted}
[
numbersep=3pt,
frame=lines,
framesep=2mm,
baselinestretch=1.2,
bgcolor=lightgraybackground,
breaklines,
linenos
]
{python}
Classifier(
  # Backbone remains frozen during HAR
  (backbone): Encoder(
    (conv1): ConvBlock(
      (conv): Conv1d(3, 32, kernel_size=(24,), stride=(1,))
      (relu): ReLU()
      (dropout): Dropout(p=0.1, inplace=False)
    )
    (conv2): ConvBlock(
      (conv): Conv1d(32, 64, kernel_size=(16,), stride=(1,))
      (relu): ReLU()
      (dropout): Dropout(p=0.1, inplace=False)
    )
    (conv3): ConvBlock(
      (conv): Conv1d(64, 96, kernel_size=(8,), stride=(1,))
      (relu): ReLU()
      (dropout): Dropout(p=0.1, inplace=False)
    )
  )
  (softmax): Sequential(
    (0): Linear(in_features=96, out_features=1024, bias=True)
    (1): ReLU(inplace=True)
    (2): Linear(in_features=1024, out_features=6, bias=True)
  )
)
\end{minted}

HAR training is performed for 50 epochs, with learning rate=$10^{-3}$, weight decay=0, and batch size=256.
Once again, we use the Adam optimizer, with the learning rate reducing by a factor of 0.8 every 10 epochs. 
The data loading and classification loops are performed using a wrapper function called \colorbox{lightgraybackground}{\texttt{evaluate\_with\_classifier()}}, to which appropriate arguments are passed. 
As the code is mostly identical to the Conv. classifier, we refer the reader to Sec.\ \ref{sec:sup_conv_classifier} for details.

\begin{minted}
[
numbersep=3pt,
frame=lines,
framesep=2mm,
baselinestretch=1.2,
bgcolor=lightgraybackground,
breaklines,
linenos
]
{python}
# Train classifier with pre-trained SimCLR encoder weights
evaluate_with_classifier(args=args)
\end{minted}

\subsection{Performance Evaluation}
\begin{figure}[h]
    \centering
    \includegraphics[width=0.45\linewidth]{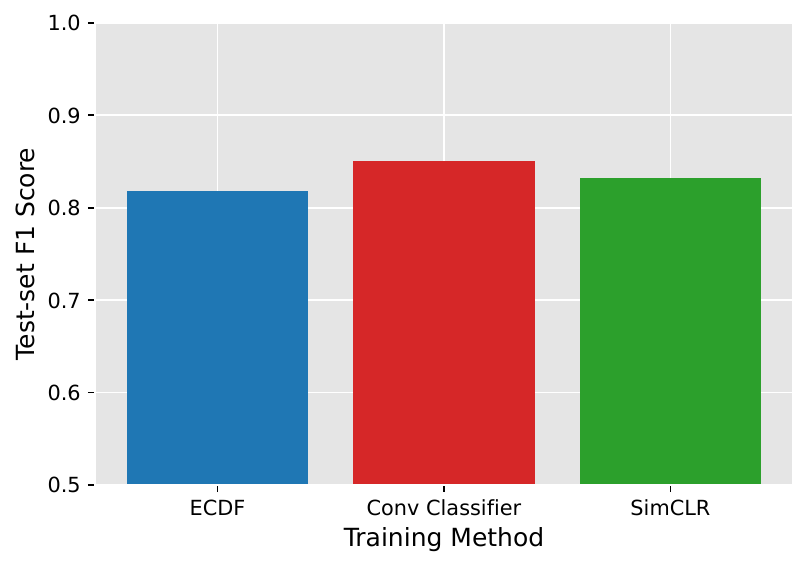}
    \caption{HAR performance obtained by the three types of representations}
    \label{fig:exp_all_perf}
\end{figure}

\begin{figure}[h]
    \centering
    \includegraphics[width=1\linewidth]{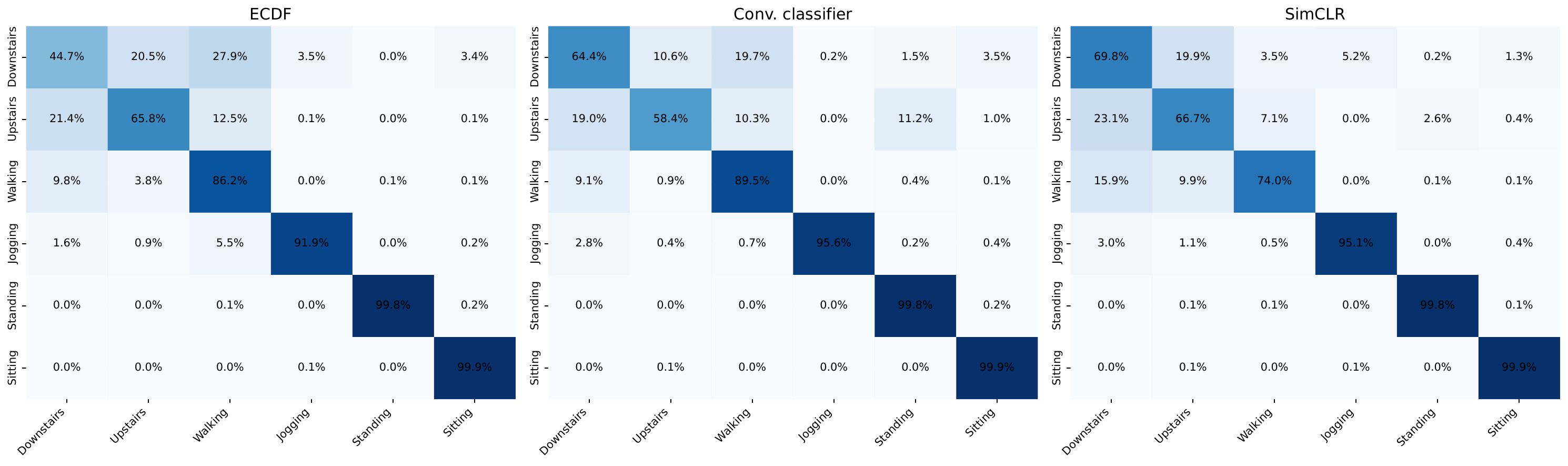}
    \caption{Confusion matrices for performing HAR using the three types of representations}
    \label{fig:exp_all_conf_matrix}
\end{figure}

Fig.\ \ref{fig:exp_all_perf} shows that end-to-end training with the Conv.\ classifier is the most effective option, whereas SimCLR is slightly worse. 
Interestingly, the ECDF-RF classifier combination is surprisingly powerful, obtaining HAR performance of 81.84\%, relative to the Conv. classifier's 85.1\%.
This showcases how ECDF features can be used for classifying simpler activities, e.g., locomotion-style activities in Motionsense, while being capable of on-the-fly extraction.

The Conv. classifier and classification using the pre-trained SimCLR encoder weights have the same architecture, albeit the learned encoder weights are frozen. 
Therefore, the number of trainable parameters is substantially lower than end-to-end training.
Yet, the performance is competitive (83.17\%), demonstrating the usefulness of self-supervised pre-training.

Comparing the confusion matrices in Sec.\ \ref{fig:exp_all_conf_matrix}, we see that the ECDF features can reliably distinguish between Jogging, Standing, and Sitting. 
Most of the confusion lies between Walking Upstairs and Walking Downstairs, and to a smaller extent, for Walking, as they have similar movements. 
Interestingly, the accuracy for Walking by the learned SimCLR representations is substantially lower than ECDF or the Conv. classifier, yet the accuracies for Walking Upstairs / Downstairs are higher, leading to better overall performance.
Meanwhile, Standing and Sitting are both static activities, but all representations are able to distinguish between them.

\end{appendix}

\end{document}